\documentclass[aps,prfluids,showpacs,showkeys,notitlepage,amsmath,amssymb,floatfix,longbibliography, 12pt, tightenlines]{revtex4-1}

\usepackage{float}
\usepackage{hyperref}
\usepackage{graphicx}
\usepackage{mathtools}
\usepackage{verbatim}
\usepackage[usenames, dvipsnames]{color}
\usepackage{natbib}
\usepackage{amsmath}
\usepackage[normalem]{ulem}

\usepackage{mathptmx}
\usepackage{hyperref}
\hypersetup{colorlinks=true, citecolor=blue, linkcolor=blue}

\newcommand{\notc}[2]{#1 \times 10^{#2}}

\begin{document}

\title{Velocity and acceleration statistics in particle-laden turbulent swirling flows}
\author{Sof\'{\i}a Angriman, Pablo D. Mininni and Pablo J. Cobelli}
\affiliation{Universidad de Buenos Aires, Facultad de Ciencias Exactas y Naturales, Departamento de F\'\i sica, \& IFIBA, CONICET, Ciudad Universitaria, Buenos Aires 1428, Argentina.}

\begin{abstract}
We present a comparison of different particles' velocity and acceleration statistics in two paradigmatic turbulent swirling flows: the von K\'arm\'an flow in a laboratory experiment, and the Taylor-Green flow in direct numerical simulations. Tracers, as well as inertial particles, are considered. Results indicate that, in spite of the differences in boundary conditions and forcing mechanisms, scaling properties and statistical quantities reveal similarities between both flows, pointing to new methods to calibrate and compare models for particles dynamics in numerical simulations, as well as to characterize the dynamics of particles in simulations and experiments. The comparison also allows us to identify contributions of the mean flow to the inertial range scaling of the particles' velocity structure functions.
\end{abstract}
\maketitle

\section{Introduction}
Turbulent flows are common in nature and industrial applications. One of their main properties is the enhancement of the mixing of quantities transported by the flow, and in recent years, significant advancements have been made in the study of turbulent particle-laden flows \citep{qureshi2007, monchaux2012, falkovich2004}. The modeling of such flows requires a wide variety of approximations, and their study in the laboratory has important consequences for flow characterization as well for practical applications. Examples of such applications include cloud dynamics and droplet formation \citep{saito2018}, aerosol and pollution dispersion in the atmosphere \citep{chandrakar2016}, and nutrient transport in the oceans \citep{abraham1998} among others \citep{delgrosso2019}. In many cases, the study of particle-laden flows has focused on the paradigmatic case of isotropic and homogeneous turbulence, a landmark in the study of turbulence. In such case, experiments of particle-laden flows are often carried out in wind tunnels, while numerical simulations use delta-correlated random forcing to sustain the turbulence against dissipation \citep{goto2008, obligado2014, monchaux2010, bec2006}.

However, in recent years significant advancements were made in the study of turbulent swirling flows in the laboratory using an experimental setup that allows generation of flows with high Reynolds numbers and with strong turbulent fluctuations superimposed on a well-defined mean flow. This includes experiments in water and air, and in different configurations (open or closed domains) and geometries (cylindrical or square cells), between two counter-rotating propellers, in a setup that generates a von K\'arm\'an flow \citep{ouellette2006, volk2011, mordant2004}. The turbulence generated using this setup is not isotropic and homogeneous, and as a result it has been sometimes called ``axisymmetric turbulence" \cite{Poncet_2008}. Experiments using this setup have been employed to study statistics of turbulence \citep{labbe1996}, turbulent transport \citep{machicoane2016, zimmermann2013}, bistabilitiy and long-term memory \citep{delatorre2007}, the emergence of singularities \citep{monchaux2006}, and even dynamo action when conducting flows are used \citep{monchaux2009}. 

Studies of tracers and inertial particles in such experiments have confirmed that turbulence in this flow displays some specific properties, and that under some circumstances transport is dominated by the strain in the center of the domain, in what have been called ``stagnation point turbulence" \citep{huck2017PRF}. Evidence of anisotropy in the flow has also been reported, as well as some common behavior with observations of particle-laden flows in isotropic and homogeneous turbulence when the mean flow is statistically removed \citep{machicoane2016}.

The von K\'arm\'an flow shares some similarities with a paradigmatic flow in the study of turbulence in periodic boundary conditions: the Taylor-Green flow \cite{taylor1937}. This flow, that displays multiple symmetries \cite{nore1997},  
has been used to study turbulence \cite{brachet1991,nore1997}, singularities of the Euler equation \cite{bustamante2012,brachet2013}, and dynamo action \cite{ponty2005}. In studies of magnetohydrodynamic dynamo action, it was successfully used to compare with von K\'arm\'an experiments, reproducing many features observed in the laboratory except for those directly associated with the different boundary conditions in both flows \citep{ponty2005, mininni2014}. It is thus worth pointing out that while some recent numerical studies consider flows with more realistic mechanical forcing and boundaries (in comparison to the von K\'arm\'an experiments) \citep{nore2018, kreuzahler2014}, the similarities between Taylor-Green and von K\'arm\'an flows still allow for interesting comparisons when it comes to attaining the largest possible Reynolds number at a fixed spatial resolution, as periodic boundary conditions are amenable to powerful and high order numerical methods.

In spite of these similarities, there are very few comparisons of Taylor-Green and von K\'arm\'an dynamics in the case of particle-laden flows. With this motivation, in this work we present a comparison of particles' velocity and acceleration statistics in these two paradigmatic turbulent swirling flows, considering on one hand laboratory experiments, and on the other direct numerical simulations. Tracers and Lagrangian particles are compared, as well as the particular case of a large inertial particle. Numerically such a particle is modeled using the Maxey-Riley equation \citep{maxey1983} in the small particle approximation, which is considered here only as an effective model. The main objective is to characterize similarities and differences between the two approaches and to evaluate the possibility of using such a comparison to validate models for particle dynamics. We consider spectra, correlation functions, single and two-times statistics, and structure functions for the particles' velocities and accelerations. In spite of the differences in boundary conditions and the forcing mechanisms, scaling properties and statistical quantities share interesting similarities between both flows, and also display a clear effect of the mean flow on particle dynamics which affect turbulent statistical properties. The comparison also allows us to disentangle contributions of the turbulence and of the mean flow to Lagrangian statistics of the particles, in particular for the often reported poor inertial range scaling of the second order Lagrangian structure function.

\section{Experimental and numerical setups}
\label{sec:setups}

\begin{figure}
  \includegraphics[trim={5.7cm 1cm 1.5cm 0},clip,width=1.\textwidth]{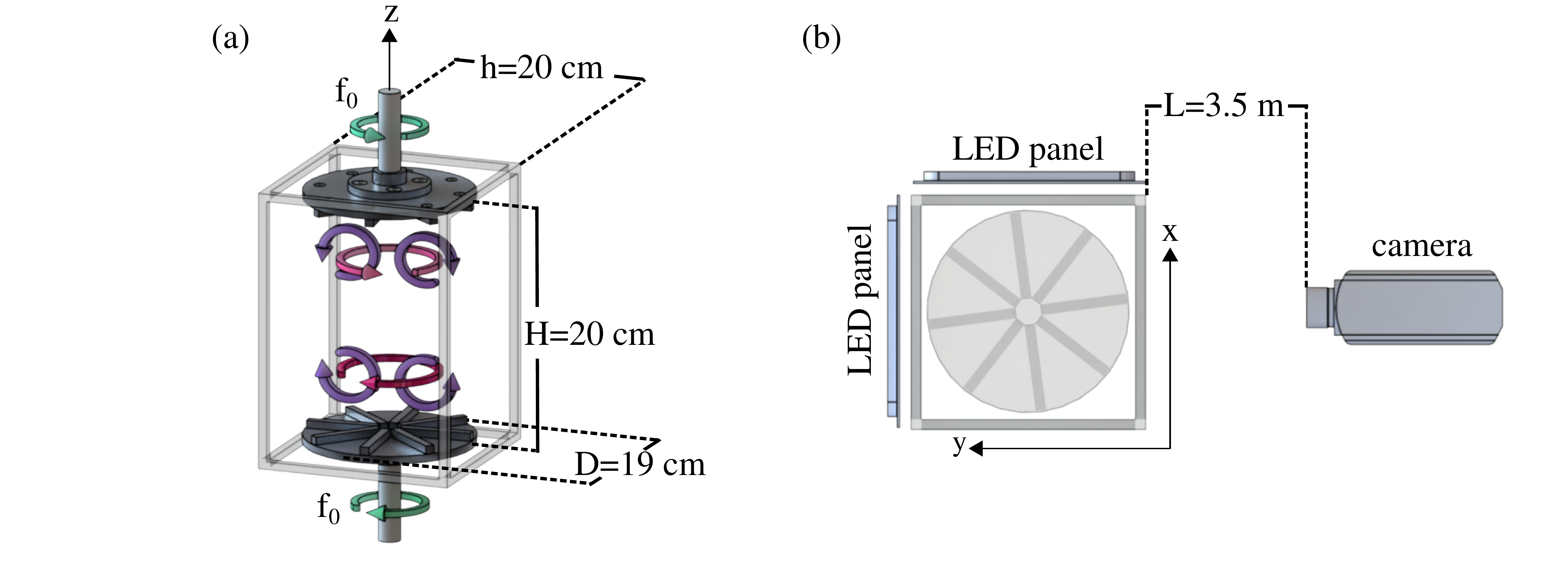}
  \caption{(a) Experimental setup, with a schematic representation of the mean large-scale flow in the von K\'arm\'an experiment. $H$ is the separation between the two impellers, $D$ is the diameter of the impellers, $h$ is the horizontal length of the cell, and $f_0$ denotes the rotation frequency of the impellers. (b) Schematic top view of the setup, with the measuring configuration (not to scale). The cell is illuminated with two LED panels, and a fast camera, placed at a distance $L$ from the cell, captures the position of the particles.}
    \label{exp_setup}
\end{figure}

\subsection{The von K\'arm\'an flow experiment \label{sec:experiment}}

The experimental setup comprises two facing disks of diameter $D=19$~cm, separated by a vertical distance of $H=20$~cm, and each fitted with 8~straight blades. The blades have a height of $1$~cm, a width of $1$~cm, and a length such that they do not reach the center of the disk (see Fig.~\ref{exp_setup}). However, their inner faces are in contact with each other in the central region of the disks, so that there is no radial flow in the surface of the disks coming from their central region. Also, there is no solid cylinder filling the central region, and thus this region is left empty. The impellers are contained in a cell of square cross-section, with side $h=20$~cm, giving access to an experimental volume of $(20\times 20 \times 20)$~cm$^3$ where the flow can be measured. The total size of the cell is $(20 \times 20 \times 50)$~cm$^3$, leaving space on the back of the impellers for shafts that connect the impellers to motors, and for refrigeration coils that allow heat removal if needed. Each impeller is driven by an independent brushless rotary servomotor (Yaskawa SMGV-20D3A61, $1.8$~kW), controlled by a servo controller (Yaskawa SGDV-8R4D01A) which provides access to the instantaneous velocity and torque of the motor. The cell is filled with distilled water from a double pass reverse osmosis system, to remove ions and dissolved or suspended solid particles from the working fluid. The setup is similar to those considered in previous laboratory studies of von K\'arm\'an flows (see, e.g., \citep{zimmermann2013}).

In all configurations considered in this study, the two disks rotate in opposite directions with angular velocity $\pm \Omega_0$, stirring the working fluid in the cell. This generates two large counter-rotating circulation cells producing, on average, a strong shear layer at the mid-plane between the disks (its detailed dynamics, however, is rather complex and it has been shown that this shear layer gets deformed and fluctuates between different configurations over time \citep{ravelet2004}). A secondary circulation in the axial direction is also generated by the impellers, resulting in a fully three-dimensional turbulent flow \citep{ravelet2008}. But as a result, the flow has a mean macroscopic structure which is anisotropic: the large-scale structures in the directions parallel to the plane of the disks are larger than the structures in the axial direction. A schematic visualization of the setup and the mean flow generated is depicted in Fig.~\ref{exp_setup}(a).

For each individual experimental run we seed the flow with either tracer or inertial particles, and stir the flow employing different values of the angular velocity $\Omega_0 = 2 \pi f_0$ (expressed in rad/s), with $f_0$ being the frequency (in Hz).  For practical reasons, we will also refer to the disks' rotational velocity as measured in revolutions per minute (rpm), which will be herein denoted by $f_0'$, with $f_0' = 60 \: f_0$.

The tracer particles are neutrally-buoyant polyethylene micro-spheres (density equal to $1$~g~cm$^{-3}$) of diameter $d = 250-300$~$\mu$m, commercially available from Cospheric. These particles are commonly used in experiments as Lagrangian tracers \citep{brown2009}. Prior to suspension, these particles were coated with a biocompatible surfactant (Tween 80) in order to ensure proper placement in suspension. For these particles we explored three different rotation velocities; namely: $f_0' = 25$, $50$~and~$100$~rpm (corresponding to $f_0 = 0.42$, $0.83$~and~$1.66$~Hz, respectively). At the largest $f_0$ considered the particles verify $d/\eta \lesssim 5$, $\eta$ being the Kolmogorov dissipation length of the flow. 

The inertial particles are $6$~mm plastic spheres with density $(0.98\pm0.01)$~g cm$^{-3}$. Particles were 3D-printed using a thermoplastic polymer (acrylonitrile butadiene styrene, or ABS). Errors in the determination of the density are mainly associated with the fact that the printed particles are slightly porous. They were injected in the flow generated by stirring exclusively at $f_0' = 50$~rpm.

In all cases, measurements of particles' dynamics are carried out using a variation of shadow particle tracking velocimetry (SPTV) \citep{huck2017_SPTV}. The cell is illuminated from two adjacent sides using two $(25 \times 25)$~cm$^2$ LED panels (each $1880$~lm, $22$~W) and a fast camera captures the particle's shadow projection over a bright background, instead of capturing reflected light as it is typically done in standard PTV techniques. Exploiting the discrete $\pi/2$ rotational symmetry of the setup about the $z$ axis, imposed by the square cross-section of the cell, only a projection of the trajectories in the $xz$-plane is registered. Individual particles are tracked in this plane using a high-speed Photron FASTCAM SA3 camera with a resolution of ($1024\times1024$)~px$^2$ and 12-bit color depth. The camera is placed in front of the cell at a distance $L=3.5$~m so that the region of observation of ($20\times20\times20$ cm$^3$) covers the whole experimental volume while warranting minimal optical distortion. A 70-300~mm lens is employed, using an effective focal length of $143$~mm. Under these experimental conditions, we estimate that the error in imaging a particle's position due to perspective effects is of $\approx 230~\mu$m, which is less than the tracers' diameter and much less than the inertial particles' diameter (see \citep{huck2017_SPTV} for other illumination techniques to reduce parallax effects).

For the tracers there are approximately $100$~particles being detected by our tracking algorithm simultaneously in each individual frame, resulting in $\mathcal{O}(10^3)$~trajectories captured after multiple realizations of the experiment, with a mean duration per trajectory of~$0.45/f_0$.  Datasets are obtained using a frame rate of $f_s = 500$~Hz, which is adequate since we are not interested in fully resolving the dissipative time scales in the experiments. For the inertial particles, in contrast, only one sphere is present in the cell in each experimental run, so as to avoid possible interactions that would result from the presence of other particles. In this case its motion within the whole experimental volume is tracked using the fast single camera but with a sampling frequency of $f_s = 125$~Hz. Multiple realizations of the experiment result in $\mathcal{O}(10^3)$~trajectories with a mean duration per trajectory of $1/f_0$. Finally, and irrespective of their nature (tracer or inertial), each particle instantaneous velocity is derived from its individual trajectory after applying a Gaussian filter.

The experiments can be characterized by two dimensionless numbers, one pertaining to the flow and another related to the particles' dynamics. An integral Reynolds number for the experiment can be defined as
\begin{equation}
    Re_\textrm{int} = \frac{2\pi f_0 ({D}/{2})^2}{\nu},
\end{equation}
where $\nu$ is the kinematic viscosity of water. The other important dimensionless parameter for the inertial particles is the Stokes number, which is usually defined as \citep{cartwright2010}
\begin{equation}
    St = \frac{2\, R^2\, (\rho_p/\rho + 1/2)}{9\, \nu}\, \frac{u}{\ell} = T_p \, \frac{u}{\ell},
    \label{eq:St}
\end{equation}
where $R$ is the particle radius, $\rho_p$ is the particle density, $\rho$ is the fluid density, $T_p$ is the particle Stokes time, and $u$ and $\ell$ are a characteristic fluid velocity and length, respectively.  Generally, $\ell$ and $u$ are chosen so that their quotient $\ell/u = \tau_\eta = (\nu/\varepsilon)^{1/2}$ is the Kolmogorov time scale of the flow; in that case we will use the notation $St_\eta$. Another possible choice is to use $\ell = L$ and $u = U$, both associated to the large (or integral) scale motion; the Stokes number resulting from this choice will be denoted as $St_\textrm{int}$. Alternatively, an effective Stokes number may be also defined as
\begin{equation}
    St_R = \frac{\tau_R}{\tau_\eta}, 
    \label{eq:StR}
\end{equation}
where $\tau_R$ is the turbulent turn over time associated with a scale $\ell$ of the order of the particle size, i.e., $\ell = R$, and therefore
\begin{equation}
    \tau_R = \left( \frac{R^2}{\varepsilon} \right)^{1/3}.
    \label{eq:tau_R}
\end{equation}
Finally, for the inertial particles we can define $Re_\textrm{p}$, the Reynolds number at the particle scale, as
\begin{equation}
	Re_\textrm{p} = \frac{R\, \lvert {\bf u} - {\bf v} \lvert }{\nu},
	\label{eq:Rep}
\end{equation}
where ${\bf u}$ and ${\bf v}$ are the flow and particle velocities, respectively.

\subsection{Taylor-Green direct numerical simulations}

We performed direct numerical simulations (herein, DNSs) of the incompressible Navier-Stokes equations
\begin{equation}
\frac{\partial {\bf u}}{\partial t} +{\bf u}\cdot{\bf \nabla}{\bf u} = -{\bf \nabla}p +\nu \nabla^{2}{\bf u}+{\bf F},
\end{equation}
where ${\bf u}$ is the solenoidal fluid velocity field ($\nabla \cdot {\bf u}=0)$, $p$ is the pressure, $\nu$ is again the kinematic viscosity, and ${\bf F}$ is an external volumetric mechanical forcing. Equations are written in dimensionless units based on a unit length $L_0$ and a unit velocity $U_0$, and solved in a three-dimensional $2\pi$-periodic cubic box using a parallel pseudo-spectral method with the GHOST code \cite{mininni2011, Rosenberg_2020}. A fixed spatial resolution of $N^3 = 768^3$ grid points is used. To mimic the geometry of the large-scale flow in the von K\'arm\'an experiments, the external forcing ${\bf F}$ is given by the Taylor-Green flow \cite{brachet1991},
\begin{equation}
    \begin{cases}
    F_x &= F_0\, \sin(k_F x)\, \cos(k_F y)\, \cos(k_F z), \\
    F_y &= -F_0\, \cos(k_F x)\, \sin(k_F y)\, \cos(k_F z), \\
    F_z &= 0,
    \end{cases}
    \label{eq:TG}
\end{equation}
with forcing wavenumber $k_F = 1$. Note that this forcing corresponds to a periodic array of counter-rotating large-scale vortices, which in the domain $[0,\pi)\times [0,\pi)\times [0,\pi)$ reduces to just two counter-rotating vortices separated vertically by a shear layer.

\begin{figure}
  {\includegraphics[width=1\textwidth]{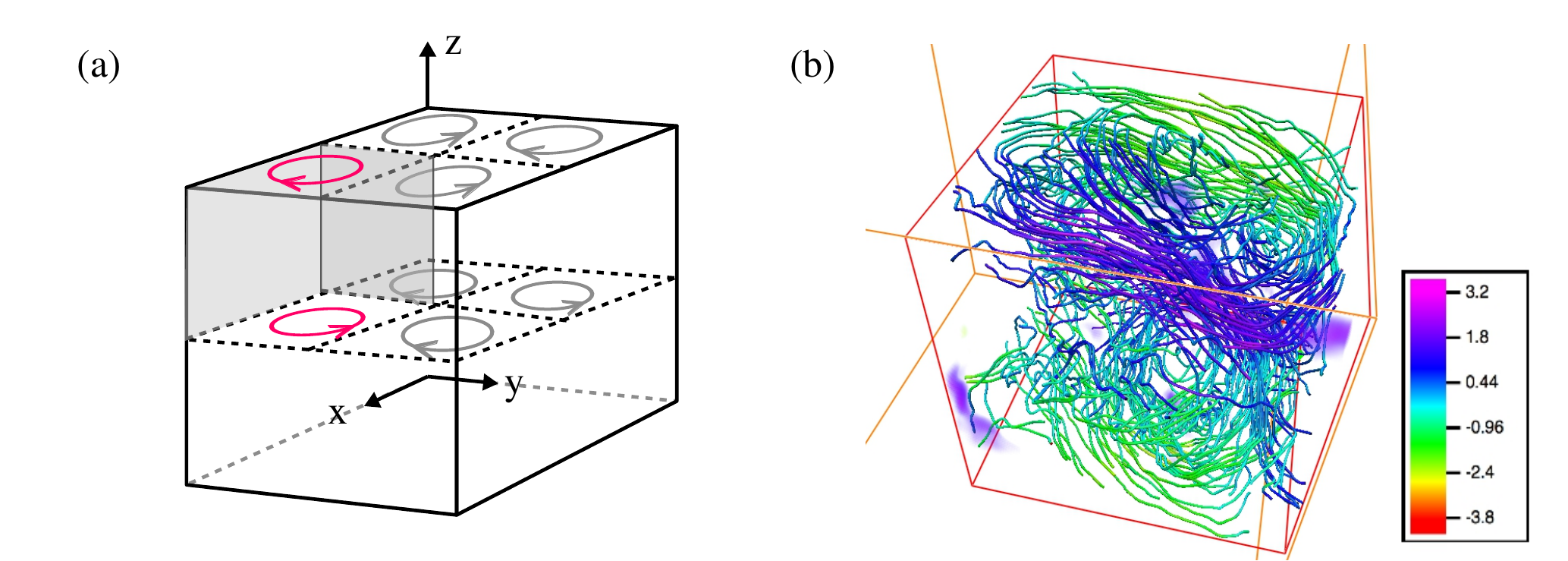}}%
  \caption{(a) Schematic of the $2\pi$-periodic 3D domain used in the DNSs. When $z$ changes in $\pi$, the Taylor-Green vortices change their sense of rotation. A region of the domain is highlighted, where the flow has a large-scale structure reminiscent of the one in the laboratory experiment (although with different boundary conditions). (b) Instantaneous streamlines in a sub-region of the solving domain. Colors denote intensity of the horizontal ($x$) component of the (Eulerian) velocity field. Note the two large-scale counter-rotating eddies.}
  \label{TG_setup}
\end{figure}

Such forcing is anisotropic: it is similar in the $x$ and $y$ directions (in fact, it has a discrete $\pi/2$ rotation symmetry about the $z$ axis as the von K\'arm\'an flow), and it injects no energy directly into the $z$ component of the velocity. As a result, the generated flow presents symmetries in a statistical sense (see \cite{brachet1991, nore1997} for more details) while keeping its anisotropy. Also as a result of these symmetries, the Taylor-Green flow in the full $(2\pi)^3$ domain can be split into 8 cells, each consisting of two counter-rotating eddies akin to those generated in the von K\'arm\'an flow as mentioned above (see Fig.~\ref{TG_setup}). As in the case of the von K\'arm\'an flow, the resulting Taylor-Green flow in this cell has a layer of strong shear in which the forcing is zero, and which separates two planes of maximum forcing. The instantaneous streamlines of the flow in a cell are shown in Figure~\ref{TG_setup}(b). As it evolves in time, the flow generates a secondary circulation in the axial direction ($\hat{z}$), driven by pressure gradients (unlike its experimental counterpart, in which Ekman pumping plays a crucial role). As mentioned in the Introduction, due to both the geometrical similarities between the Taylor-Green and von K\'arm\'an flows and the convenience of having a similar flow with periodic boundary conditions for the numerical study of turbulence, this forcing has been used in many cases to compare simulations with experimental data obtained from von K\'arm\'an setups \citep{krstulovic2011, mininni2014}, specially for magnetohydrodynamic dynamo studies.

In our DNSs, the flow is first evolved until a turbulent steady state is reached. From that instant on, point particles are injected and evolved in time together with the flow, while computing their instantaneous position, velocity, and acceleration. Particles do not interact with each other, and their dynamics do not affect the flow evolution. Multiple simulations with different particles are then performed, in each case with $10^6$~particles.

We study the dynamics of two types of particles. Firstly, we consider Lagrangian tracers which evolve according to
\begin{equation}
    \frac{d {\bf x}_p}{dt} = {\bf u}({\bf x}_p,t),
    \label{eq:tracers_pc}
\end{equation}
where ${\bf x}_p(t)$ is the position of the Lagrangian tracer at time $t$, and ${\bf u}({\bf x}_p,t)$ is the velocity of the fluid element at position ${\bf x}_p(t)$.

Secondly, the evolution of inertial particles will be described by an effective simple model based on the equations for the dynamics of inertial neutrally-buoyant point particles. In principle, sufficiently small inertial particles with a density mismatch to that of the fluid can modeled by the Maxey-Riley equation \citep{maxey1983, cartwright2010}, which when considering only first order effects in the particle radius reads
\begin{equation}
    \frac{d {\bf v}}{dt} =  \frac{1}{T_p} \left[{\bf u}({\bf x}_p,t) - {\bf v}(t)\right] + \frac{3}{2}\frac{\gamma}{1+\gamma/2} \frac{D {\bf u}({\bf x}_p,t)}{Dt} + \frac{1-\gamma}{1+\gamma/2}{\bf g},
    \label{eq:MR}
\end{equation}
where ${\bf v}(t)$ is the particle's velocity, ${\bf u}({\bf x}_p,t)$ is again the fluid velocity at the particle position, $T_p$ is the Stokes time defined as in Eq.~(\ref{eq:St}), $\gamma=\rho/\rho_p$ is the fluid to particle density ratio, $D/Dt$ is the convective derivative, and ${\bf g}$ is the acceleration of gravity. This approximation is valid for particles with a radius $R$ much smaller than all the characteristic lengths of the flow, and provided the velocity difference between the particle and the flow is not large (i.e., provided $Re_\textrm{p} \ll 1$). The equation, neglecting the effect of gravity, has been previously used to study the dynamics of inertial particles in turbulent flows using DNSs \citep{Calzavarini2008}, and to compare the particles' acceleration statistics with data obtained from von K\'arm\'an experiments \citep{Volk2008}. However, no simple equation is available for particles with size much larger than the Kolmogorov scale $\eta$, even if higher order corrections in $R$ are taken into account in Eq.~(\ref{eq:MR}). As a result, we will use a much simpler equation, which we will evaluate as a phenomenological model with an effective coefficient. We consider just the effect of Stokes drag, assuming the velocity of large particles tries to match the fluid velocity with an effective Stokes relaxation time $\tau_p$, with two main goals: (1) To see what features in the experiments can be modeled by such a simple equation, and (2) to identify ways of computing an effective Stokes time such that results from the experiments are reasonably well captured by the DNSs. The equation for the evolution of the inertial particles in our DNSs is then given by
\begin{equation}
    \frac{d {\bf x}_p}{dt} = {\bf v}(t) , \qquad
    \frac{d {\bf v}}{dt} =  \frac{1}{\tau_p} \left[{\bf u}({\bf x}_p,t) - {\bf v}(t)\right].
    \label{eq:iner}
\end{equation}

Both for the tracers and for the inertial particles, integration of the corresponding dynamical equations (either Eq.\eqref{eq:tracers_pc} or Eq.~\eqref{eq:iner}) is performed using a high-order Runge-Kutta time stepping scheme combined with a high-order three-dimensional spatial spline interpolation to obtain the fluid velocity ${\bf u}({\bf x}_p,t)$ at the position of the particles \cite{Yeung_1988}.

Just as in the experiments, for these ``effective'' inertial particles the Stokes number $St$ is defined as the ratio of two characteristic times: the relaxation time of the particle, $\tau_p$, and some characteristic time of the fluid $\tau_\ell$ at scale $\ell$, so that  $St = \tau_p/\tau_\ell$. As in Sec.~\ref{sec:experiment}, when $\tau_\ell$ is the the Kolmogorov dissipation time scale $\tau_\eta$, we obtain $St_\eta = \tau_p/\tau_\eta$. When $\tau_\ell$ is evaluated at the flow integral scale $L$, we obtain $St_\textrm{int} = \tau_p/\tau_L$. And finally, it is worth noting that Eq.~\eqref{eq:tau_R} allows us to estimate an effective radius $R$ for the (otherwise point) inertial particles in the simulations, as
\begin{equation}
    R = \tau_p^{3/2} \varepsilon^{1/2}.
\end{equation}
From this relation, we can also compute an effective Stokes number $St_R$ using Eq.~(\ref{eq:StR}), and the Reynolds number based on the particle scale $Re_\textrm{p}$ using Eq.~(\ref{eq:Rep}).

\section{Lagrangian tracers}
\label{sect:lag-tracers}

\begin{table}
\begin{ruledtabular}
\begin{tabular}{l c c c c c c c c c c}
\hfill
    Dataset & $f_0'$ & U & L  & $\tau_L^{(x)} f_0$& $\varepsilon_L$  &$\eta$ & $Re_\textrm{int}$ & $Re_\textrm{part}$ & $R_{\lambda}$ & $\tau_L^{(x)}/\tau_L^{(z)}$ \\
    \hline
            & [rpm] & [m/s] & [m] & & [W/kg] & [$\mu$m]  &  &  & \\
    EXP25   & 25    & 0.033 & 0.19   & $0.36$&  $\notc{6}{-4}$   & 200 & $\notc{2.4}{4}$ & $\notc{0.6}{4}$ & $170$ &  $1.27$\\
    EXP50   & 50    & 0.083 & 0.19   & $0.34$&  $\notc{8.5}{-3}$ & 105 & $\notc{4.7}{4}$ & $\notc{1.6}{4}$ & $290$ &  $1.39$\\
    EXP100  & 100   & 0.168 & 0.19   & $0.33$&  $\notc{7.1}{-2}$ & 60  & $\notc{9.5}{4}$ & $\notc{3.2}{4}$ & $410$ &  $1.41$\\
    \hline
    DNS     &  $1/2\pi$ & 0.904  & $2\pi$  & $0.35$&  $\notc{2.4}{-1}$& $\notc{4.4}{-3}$ &- & $\notc{1.3}{4}$ & $305$ & $1.27$        
\end{tabular}
\end{ruledtabular}
\caption{Values of the parameters for both experiments and simulations with tracers. DNS values are dimensionless. For the experiments, $f'_0$ corresponds to the rpm frequency of the disks ($f'_0 = 60 \: f_0$, where $f_0$ is the frequency in s$^{-1}$). For the DNS, $f_0$ is the frequency associated to a large-scale eddy turn over time. $U$ is the r.m.s.~value of $v_x$, the $x$ component of the particles' velocity, $L$ is the flow integral scale, and $\tau_L^{(x)}$ is the particle velocity autocorrelation time based on $v_x$. The energy injection rate is given by $\varepsilon_L = U^2/(2 \tau_L^{(x)})$ in the experiments, and measured directly from the injected power in the simulations. The Kolmogorov dissipation scale is $\eta = (\nu^3/\varepsilon_L)^{1/4}$. $Re_\textrm{int}$ and $Re_\textrm{part}$ are respectively the integral and tracer-based Reynolds numbers. The Taylor-based Reynolds number is $R_\lambda = \sqrt{15\, U^4/\nu\,\varepsilon_L}$, and $\tau_L^{(x)}/\tau_L^{(z)}$ is the ratio of the particles' autocorrelation times based on $v_x$ and $v_z$.}
\label{Lag_table}
\end{table}

To study the dynamics of the von K\'arm\'an flow from a Lagrangian viewpoint, we track the evolution of the tracers (described in Sec. \ref{sec:experiment}) using three different rotation velocities: $f'_0 = 25,\, 50$ and $100$ rpm, following their trajectories in the horizontal ($x$) and axial ($z$) directions. For reference, these datasets will be labeled EXP25, EXP50, and EXP100 respectively (see Table \ref{Lag_table}). In the Taylor-Green numerical simulations, Lagrangian characterization is done by evolving tracers according to Eq.~ (\ref{eq:tracers_pc}). The measurements of the tracers' velocities, together with the Lagrangian autocorrelation times $\tau_L^{(i)}$ (computed as the first zero-crossing of the autocorrelation function of the $i$-th component of the particle's velocity, as detailed in Sec.~\ref{sec:lag_corr}), lead us to define quantities that supplement those defined in Sec.~\ref{sec:setups}, and that will allow for comparisons of experiments and simulations on equal footing. 

We start with the quantification of the energy injection rate. The numerical simulations provide us with direct access to the energy injection rate $\varepsilon$ from the computation of the power injected by the forcing. In contrast, in the experiments we will estimate the energy injection rate from the r.m.s.~value of the horizontal component of the particles' velocity, $v_x$, denoted herein by $U$, and from the time $\tau_L^{(x)}$, the Lagrangian autocorrelation time of $v_x$, as
\begin{equation}
    \varepsilon_L = \frac{1}{2} \frac{U^2}{\tau_L^{(x)}}.
    \label{eq:eps}
\end{equation}
The choice of using the $x$ component of the velocity in this definition results from the fact that, both in the numerics and in the experimental flow, the largest eddy lies in this direction. In addition, for all the datasets considered, the tracers decorrelate in a similar fashion in this direction. Moreover, the estimation of the energy injection rate from Eq.~(\ref{eq:eps}) scales as expected with increasing $f_0$. On the contrary, usual estimates of the energy injection rate measured directly from the power consumption of the motors $\varepsilon_{\textrm{pow}}$ (as done, e.g., in \citep{machicoane2014, kuzzay2015}) do not scale as expected with increasing $f_0$ for our experiments. This behavior could be a consequence of moderate Reynolds number effects (arising, e.g., from significant energy losses associated with friction and boundary layer effects), but could also be attributed to specific properties of von K\'arm\'an flows in square cells, such as those discussed in \citep{huck2017PRF}. Most certainly, the former effects result from the fact that, in order to facilitate comparisons with numerical simulations, we employ values of $f_0$ lying on the lower range of those considered in previous studies of von K\'arm\'an flows. The latter effects result from the choice of using a square cell, which also facilitates comparisons with the simulations. It is also worth mentioning that estimating the energy injection rate from the zero crossing of the acceleration autocorrelation function, as done in \citep{huck2017PRF}, yields values similar to those obtained from our definition of $\varepsilon_L$ in Eq.~\eqref{eq:eps}. Still, it should be noted that to our knowledge, the definition used here has not been used in the past by other groups doing experiments of von K\'arm\'an flows. Table \ref{Lag_table} lists all values of $\varepsilon_L$ for the experiments, while for the simulation it lists $\varepsilon_L=\varepsilon$.

Using the r.m.s.~velocity $U$ we can now define a Reynolds number based on the particles' velocity (not to be confused with $Re_\textrm{p}$) as
\begin{equation}
    Re_\textrm{part} = \frac{U\, L}{\nu},
    \label{eq:Repart}
\end{equation}
where $L$ is associated to the forcing scale and is provided in Table \ref{Lag_table}. For the simulations, $Re_\textrm{part} \approx Re_\textrm{int}$. Using $U$ and $\varepsilon_L$, the Taylor-based Reynolds number can now be also defined as
\begin{equation}
    Re_\lambda = \sqrt{\frac{15 U^4}{\nu \varepsilon_L}}.
    \label{eq:Relamb}
\end{equation}

All the relevant parameters for the experimental and numerical datasets are listed in Table~\ref{Lag_table}. Note that from the values of $Re_\textrm{part}$ and $Re_\lambda$, EXP50 (i.e., the experiment with $f'_0 = 50$~rpm) and the DNS share comparable values of the Reynolds number (note other dimensionless numbers are also comparable between this experiment and the simulation). In the following, and in light of these similarities, comparisons between experiments and simulations for Lagrangian tracers will focus on these two cases.

\begin{figure}
    \includegraphics[width=1\textwidth]{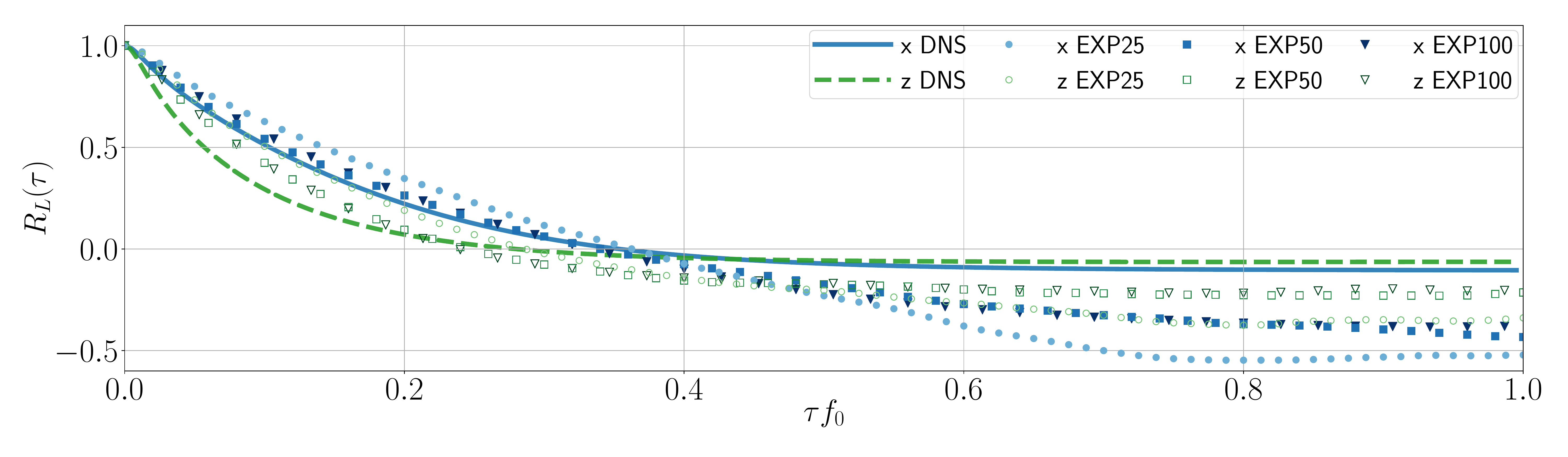}
    \caption{Particle velocity autocorrelation function for the horizontal ($x$) and axial ($z$) components of the velocity, in experiments and in simulations. The time axis is normalized by $f_0$ (the rotation frequency of the blades in the experiments, and the frequency associated to the largest eddy turnover time in the DNSs).}
    \label{lag:corr}
  \end{figure}

\subsection{Velocity autocorrelation functions \label{sec:lag_corr}}

We start by characterizing tracers' dynamics by means of the Lagrangian autocorrelation function of particles' velocities. For a single Cartesian component of the tracer velocity $v_i$, the  Lagrangian (normalized) autocorrelation function is given by
\begin{equation}
    R_L^{(i)}(\tau) \equiv \frac{C_v(\tau)}{C_v(0)} =  \frac{\langle v_i(t) v_i(t+\tau) \rangle}{\langle v_i^2(t) \rangle}  
    \label{eq:corr}
\end{equation}
where the brackets $\langle \cdot \rangle$ denote averages over the time $t$ and over all trajectories, and where $\tau$ is the time lag. To compute this magnitude, only experimental velocity tracks with a duration longer than $1/f_0$ were included in the ensemble averaging, so as to capture contributions from the mean flow, and to maintain consistency in the comparisons with data stemming from the DNS (in which all particles can be tracked for arbitrarily long times). No significant bias was observed by doing so. The resulting autocorrelation functions for both the experiments and the simulations are shown in Fig.~\ref{lag:corr}, for the horizontal ($i=x$) and axial ($i=z$) velocity components.

For the sake of clarity and consistency in the graphical representation of our results, we shall adopt the following convention throughout the rest of the paper. Symbols (lines) are used to identify experimental (numerical) results. Full symbols or continuous lines (depending on whether the data is experimental or numerical) denote the horizontal component ($x$), whereas empty symbols or dashed lines represent the axial component ($z$). In the case of the experiments, runs performed at $f'_0 = 25, 50$, and $100$~rpm are symbolized by circles, squares, and upside triangles respectively.   

Firstly, it is worth noting that, both in the experiments and in the simulations, the Lagrangian autocorrelation function becomes negative for long times and does not converge rapidly to zero afterwards. This can be interpreted as an effect of the mean flow in the system. Although the effects of the mean flow in the statistics can be partially alleviated by studying particles' statistics in a smaller subdomain of the von K\'arm\'an cell or by carefully removing mean flow components (see, e.g., \citep{machicoane2016, huck2019}), in our case we wish to compare quantities in simulations and experiments without removing the effects associated with the large-scale flow, so as to identify global similarities and differences between both setups. Secondly, we observe that for short times the autocorrelation of the axial velocity component, $R_L^{(z)}$, decays faster with $\tau$ than its horizontal counterpart, $R_L^{(x)}$, and that this behavior is also common to both the experiments (for all values of the Reynolds number considered in this study) and the simulation.

From Fig.~\ref{lag:corr} we can estimate $\tau_L^{(i)}$, the tracers' component-wise velocity autocorrelation time, as the instant corresponding to the first zero-crossing of the corresponding autocorrelation function. The differences in the decay of $R_L^{(x)}$ and $R_L^{(z)}$ with $\tau$ show that the ratio $\tau_L^{(x)}/\tau_L^{(z)} > 1$, which is a signature of the flow anisotropy and of the effect of the large-scale circulation. This ratio grows as the Reynolds number is increased (see values in the right-most column of Table \ref{Lag_table}). However, the product $f_0 \tau_L^{(x)}$ is comparable for all datasets, numerical or experimental. This further motivates using $\tau_L^{(x)}$  in the computation of the energy injection rate $\varepsilon_L$ in Eq.~(\ref{eq:eps}). A related point to consider is the fact that the autocorrelation functions for $f_0' = 50$~rpm and $f_0' = 100$~rpm (i.e., those with the highest values of the Reynolds numbers considered) collapse component-wise for nearly all time lags. This could indicate that the main contributor to the decorrelation of particles' velocities is the mean flow, which is expected to vary less as the Reynolds number becomes sufficiently large. This is also reinforced by the fact that the data corresponding to $f'_0 = 25$~rpm, while still displaying a similar qualitative behavior to the other experimental datasets, do not collapse with the other curves, which is consistent with the fact that the Reynolds number in this experiment is close to the threshold for developed turbulence in von K\'arm\'an setups \citep{ravelet2008, cortet2009}.

\subsection{Velocity power spectra \label{sec:lag_spec}}

\begin{figure}
  {\includegraphics[width=0.5\textwidth]{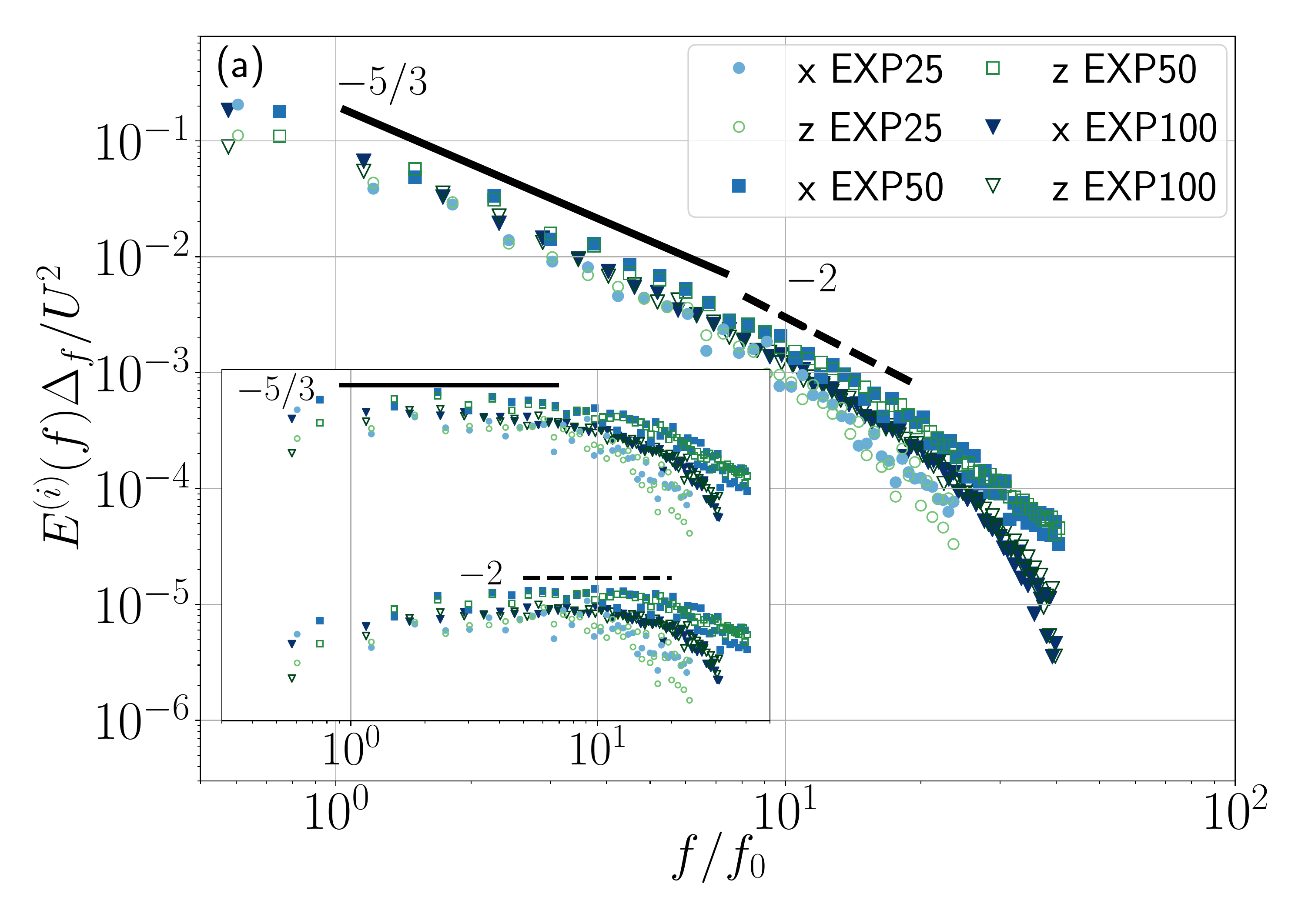}}%
  \hfill
  {\includegraphics[width=0.5\textwidth]{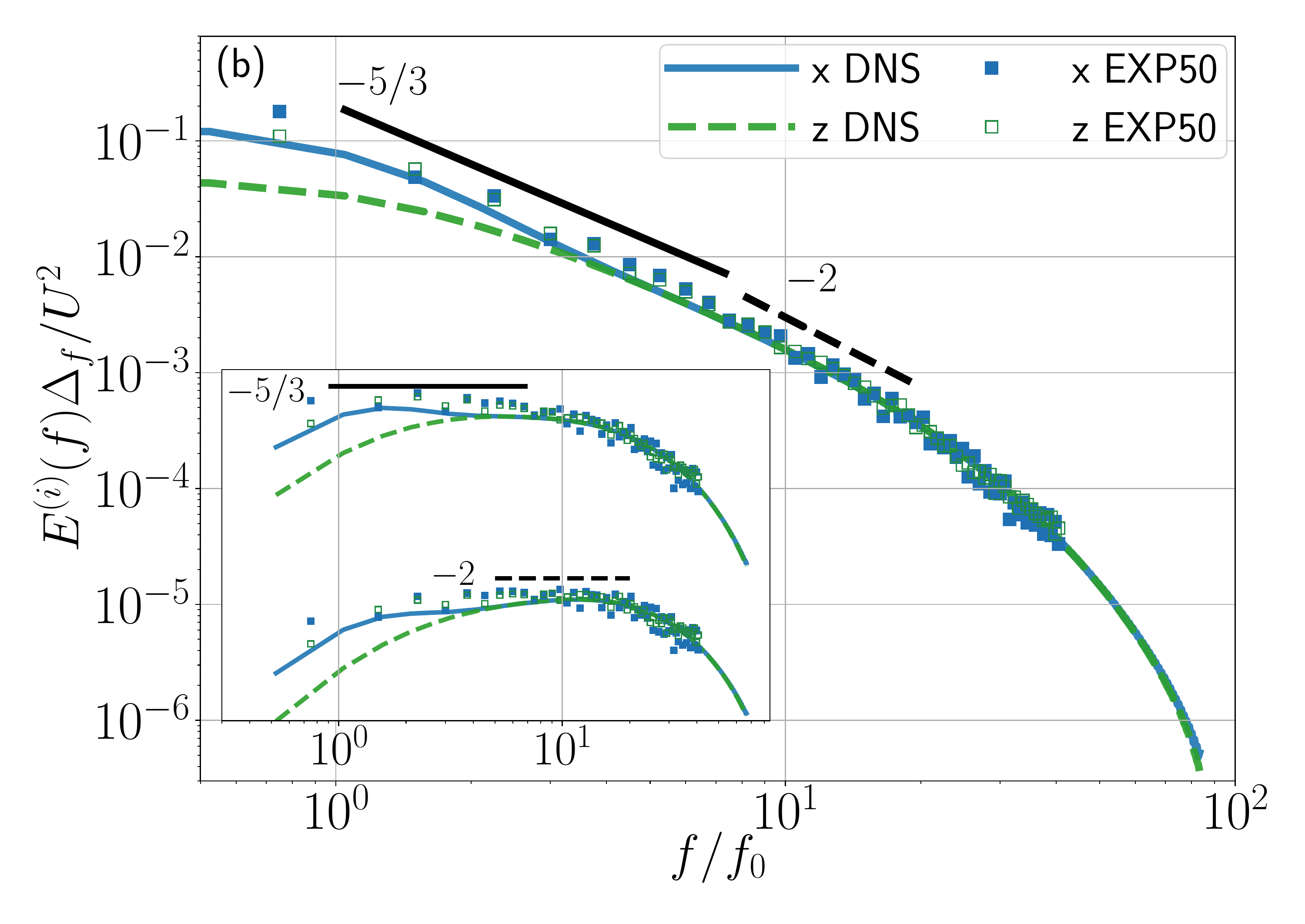}}
  \caption{Lagrangian velocity power spectra in log-log scale. (a) Experimental results, for the different rotation frequencies $f_0$, and for the different velocity components. The inset shows the same spectra compensated by power laws with exponents $-5/3$ (top) and $-2$ (bottom). The spectra in the inset with the two compensations were separated by multiplying them by an arbitrary factor for the sake of clarity. (b) Comparison of the DNS Lagrangian spectrum with that from EXP50, the experimental dataset with the closest matching value for the Reynolds number. The inset shows the compensated spectra with the same conventions.}
  \label{lag:energy_spec}
\end{figure}

The power spectra of the tracers' velocities for the horizontal and axial velocity components in the three experiments and in the DNS are depicted in Fig.~\ref{lag:energy_spec}. For the experiments, Fig.~\ref{lag:energy_spec}(a) shows that spectra are compatible with a scaling law over a frequency range exceeding a decade. For low frequencies (i.e., long time scales associated with large scale motions) the power law exponent is close to $-5/3$ (see the inset for compensated spectra). For intermediate frequencies, the power law may be compatible with a $-2$ scaling, albeit in a shorter range of wavenumbers (see also the inset). This behavior is somehow unexpected: a Kolmogorov scaling of the Eulerian energy spectrum $E(k) \propto k^{-5/3}$ is expected to yield a Lagrangian spectrum $E(f) \propto f^{-2}$, as Lagrangian trajectories are not expected to be affected by sweeping (see, for example, \citep{Lvov1997,ToschiBodenschatz} and references therein, and the discussion in Sec.~\ref{sec:S2}) which results in the $\propto f^{-5/3}$ Eulerian frequency spectrum. As will be confirmed by the second order Lagrangian structure functions and the acceleration spectrum, our data indicates that sweeping by the large-scale flow plays a relevant role in the particle evolution even in the Lagrangian frame.

In Fig.~\ref{lag:energy_spec}(b) we compare the power spectra of the tracer's velocities from EXP50 (the experiment with $f_0 = 50$ rpm) with those resulting from the simulations. As mentioned before, these datasets have the closest matching values of $Re_\textrm{part}$ and $Re_\lambda$. All curves collapse for nearly all time scales (including forcing and dissipative time scales), and scaling ranges compatible with both power laws ($-5/3$ and $-2$) are again identifiable in both datasets. The DNS spectra are slightly more anisotropic at the largest scales, as the curves for the $x$ and $z$ velocity components have a larger relative difference in their amplitudes. Indeed, the ratio $U/U_z$, where $U$ ($U_z$) is the r.m.s.~value of the $x$ ($z$) component of the tracers' velocity, is larger for the DNS than for EXP50: for the DNS $U^{\textrm{DNS}}/U_z^{\text{DNS}} \approx 1.4$, while for EXP50 $U^{\textrm{EXP50}}/U_z^{\textrm{EXP50}} \approx 1.2$ (this ratio has similar values for all the other experimental datasets). Except for this difference, the power spectra of the tracers' velocities in the experiment at $f'_0=50$~rpm and the simulation show good agreement. Note also that this quantification of the anisotropy, while slightly smaller than that reported in other von K\'arm\'an experiments in square cells \cite{huck2017PRF}, is also consistent with the fact that we use a cubic cell with unitary aspect ratio (i.e., disks are separated by a distance equal to the cell of 20 cm width and depth), while results in e.g., Ref.~\cite{huck2017PRF}, have the ratio between these two lengths equal to 4:3.

\subsection{Structure functions \label{sec:S2}}

\begin{figure}
  {\includegraphics[width=0.5\textwidth]{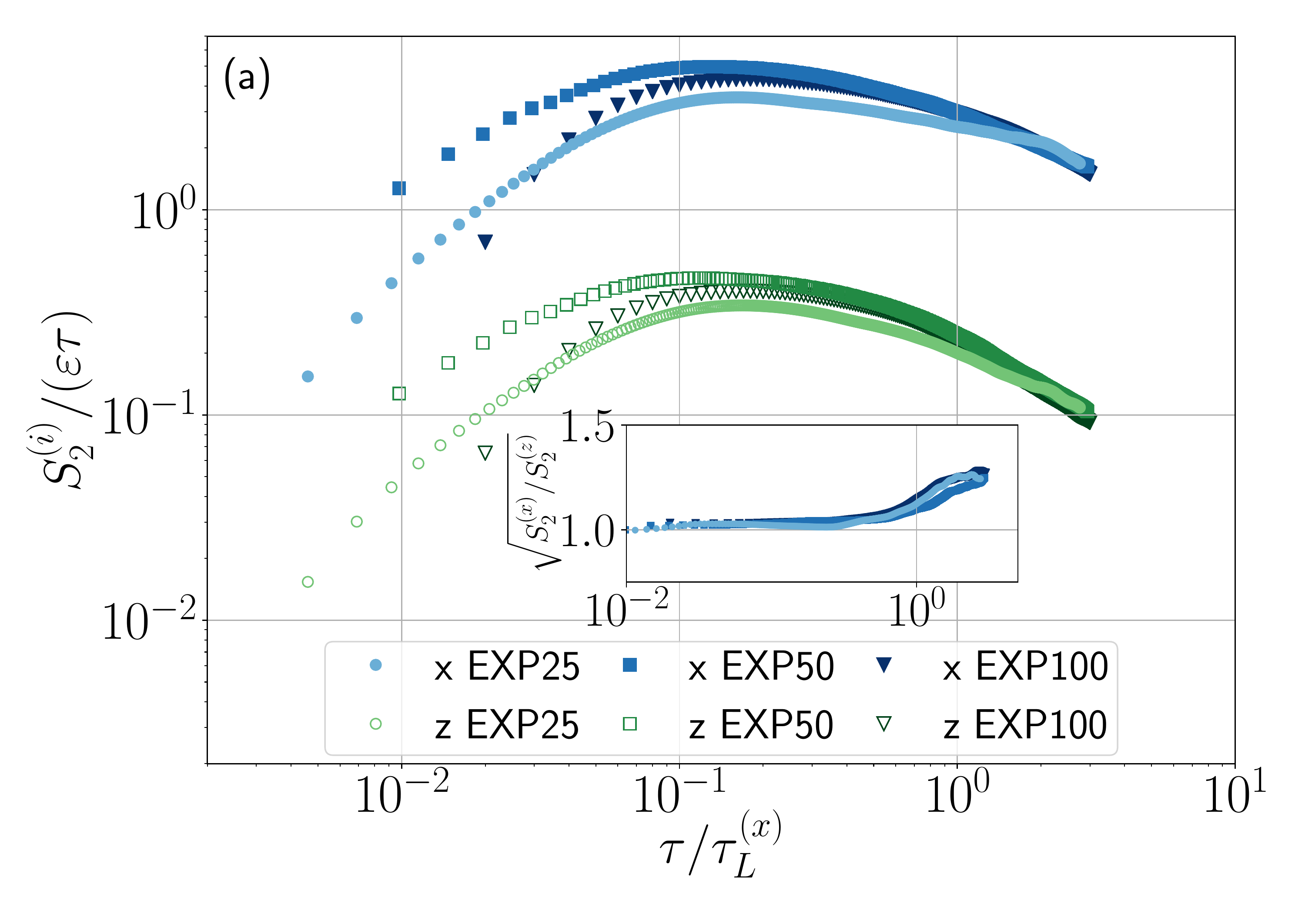}}%
  \hfill
  {\includegraphics[width=0.5\textwidth]{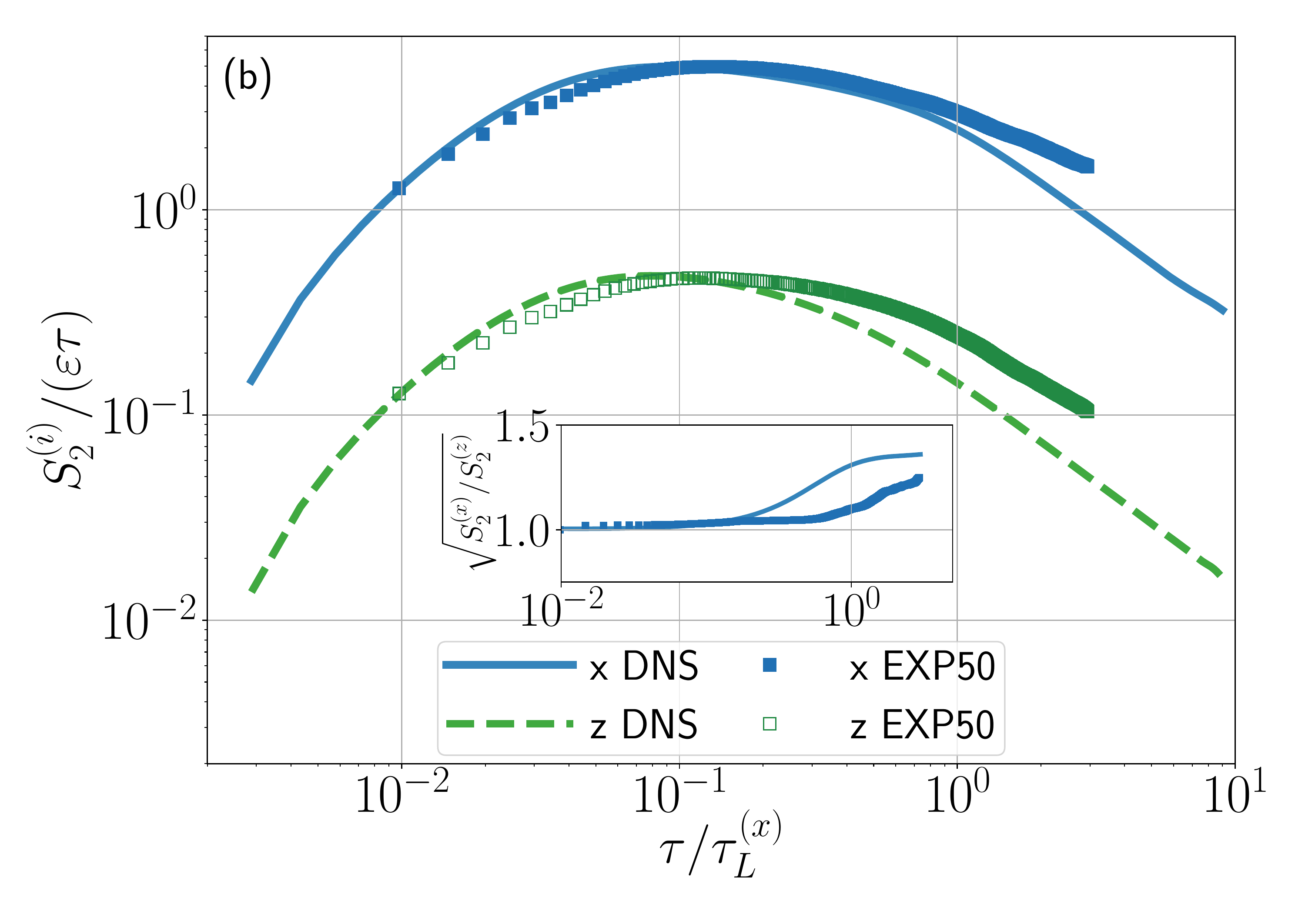}}
  \caption{Second order tracers' velocity structure functions (for each velocity component separately), compensated by the inertial range prediction $\varepsilon \tau$, for: (a) Experimental data, and (b) DNS data compared with EXP50. In both figures the curves for the $z$ component of the velocity are shown with an arbitrary vertical displacement of $10^{-1}$. The insets show $\sqrt(S_2^{(x)}/S_2^{(z)})$ using linear scale in the vertical axis, for the three experiments in panel (a), and for EXP50 and the DNS in panel (b).}
  \label{lag:S2_tau}
\end{figure}

The tracers' (component-wise) second order velocity structure function is given by
\begin{equation}
    S_2^{(i)} (\tau) = \langle [v_i (t+\tau) - v_i(t)]^2 \rangle,
    \label{eq:S2}
\end{equation}
where again the index $i$ denotes the Cartesian component of the velocity considered, and $\tau$ is the time lag. In Fig.~\ref{lag:S2_tau} we show these structure functions for each component, both for the experiments and the DNS. While for very small time lags we can expect $S_2^{(i)} \propto \tau^2$ from the regularity of the velocity field, for intermediate time lags (i.e., in the turbulent inertial range) the prediction for isotropic and homogeneous turbulence is $S_2^{(i)} \sim \varepsilon \tau$ \cite{monin1975}. Such a behavior is compatible with the prediction for the Lagrangian energy spectrum $E(f) \propto f^{-2}$. As a result, in Fig.~\ref{lag:S2_tau} the structure functions are compensated by $\varepsilon \tau$. A very short range with constant $S_2^{(i)}/(\varepsilon \tau )$ is seen, slightly more clearly for EXP100 with $f'_0=100$ rpm. The lack of clear scaling has already been pointed out in the literature, for both numerical and experimental data \citep{mordant2001, xu2006, biferale2008}. It has been reported that the reason for this can be that this quantity mixes low-frequency and inertial-range fluctuations \citep{huang2013}, and that it converges very slowly towards its asymptotic value reaching a plateau only for $Re_\lambda \gtrsim \notc{3}{4}$ \citep{sawford2011}.

The structure functions provide us with an alternative way of quantifying the anisotropy in the tracers' r.m.s.~velocity. To this end we compute $\sqrt(S_2^{(x)}/S_2^{(z)})$, shown in the insets in Fig.~\ref{lag:S2_tau}. On the one hand, for values of $\tau/\tau_L^{(x)} \approx 1$ or slightly smaller this ratio is comparable to the values of $U/U_z$ discussed in Sec.~\ref{sec:lag_spec}, and it becomes larger for larger time scales (i.e., for $\tau>\tau_L^{(x)}$). On the other hand, for time lags $\tau/\tau_L^{(x)} \lesssim 10^{-1}$ the ratio approaches a value of $1$ both in the experiments and in the simulations. This again confirms that the large-scale von K\'arm\'an flow generates anistropies in the fluctuations of the velocity, while the slightly smaller amount of anisotropy when compared with previous studies as those reported in \citet{huck2017PRF} could be associated with the differences in the aspect ratio of our experimental cell.

Another effect associated to the mean flow (which has low-frequency components), both in the von K\'arm\'an and in the Taylor-Green flows, can be the observed anomaly in the expected $S_2^{(i)} \sim \varepsilon \tau$ scaling of the Lagrangian structure function. Indeed, the mean flow can introduce a decorrelation time associated with a large-scale turnover time, resulting from the sweeping of the tracers by the largest eddies. This argument, which is often considered in the Eulerian frame \cite{Tennekes_1975, ChenKraichnan}, may be also important in Lagrangian measurements as indicated by recent models of Lagrangian dispersion \citep{sujovolsky2018, rast2011}.
Kraichnan already noted this when developing the Lagrangian-History Direct Interaction Approximation (LHDIA), and introduced mixed Eulerian-Lagrangian correlations (based on his so-called ``generalized velocity'') to fully remove the advection of eddies by a large-scale flow \citep{Kraichnan1965, *Kraichnan1966errata, Kraichnan1966}. In a similar bridging effort between the Eulerian and Lagrangian descriptions, \citet{BelinicherLvov1987} showed that sweeping effects can be eliminated by using a reference frame (termed the quasi-Lagrangian reference frame) shared by all fluid points inside a large eddy (see also \citep{LvovProcaccia1995,LvovProcaccia1996}, and references therein). 

\begin{figure}
  {\includegraphics[width=0.5\textwidth]{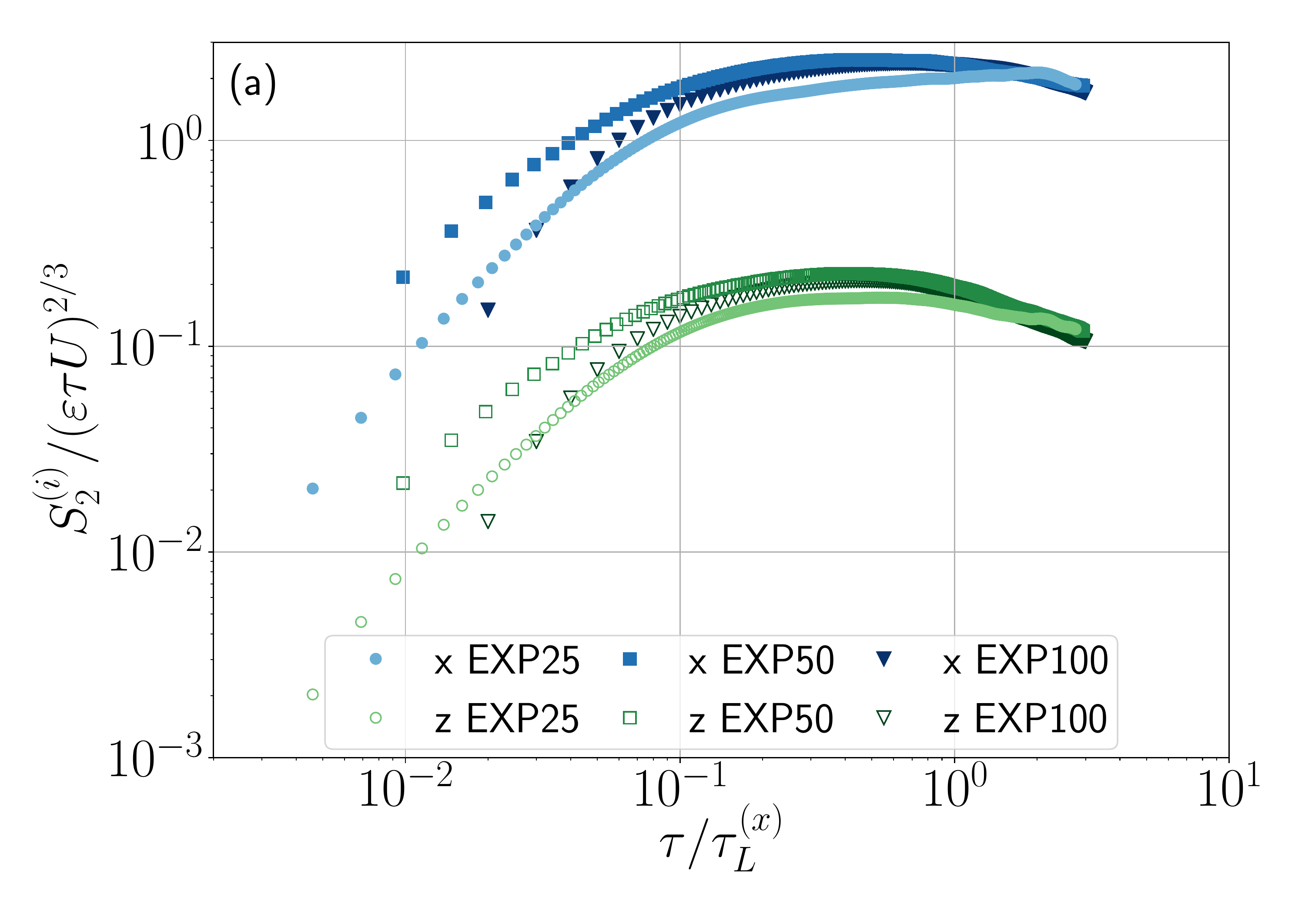}}%
  \hfill%
  {\includegraphics[width=0.5\textwidth]{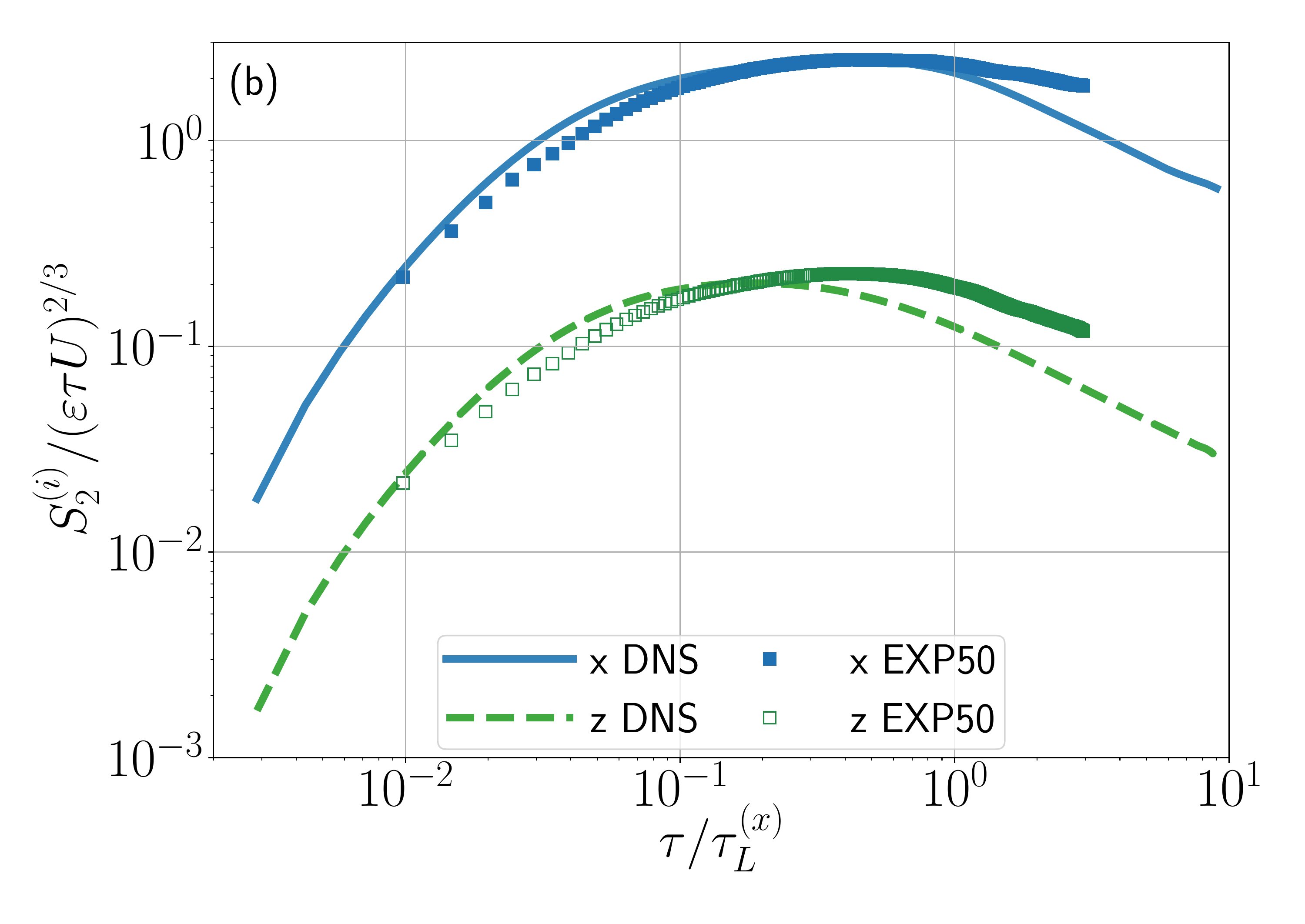}}
  \caption{Second order tracers' velocity structure function as in Fig.~\ref{lag:S2_tau}, but compensated by a sweeping-dominated prediction $(\varepsilon\tau\, U)^{2/3}$, for: (a) Experimental data, and (b) DNS data compared with EXP50. In both figures the curves for the $z$ component of the velocity are shown with an arbitrary vertical displacement by multiplying them by $10^{-1}$.}
  \label{lag:S2_tau23}
\end{figure}

If sweeping by the large-scale eddies is indeed affecting the second order Lagrangian statistics, we can estimate its effect in the context of Kolmogorov's theory. In the inertial range, the second order Eulerian structure function scales as $S_2(\ell) \sim (\varepsilon\, \ell)^{2/3}$. If instead of assuming that decorrelation in the measurements is controlled by the local turnover time we consider the decorrelation from the large-scale eddies, then $\tau \sim \ell/U$ instead of $\tau \sim \ell/u_\ell$ (where $U$ is again the r.m.s.~value of the total velocity of the flow, taking into account the contributions of the mean flow and of the turbulent fluctuations, and $u_\ell$ is the characteristic velocity at scale $\ell$). As a result, it follows that
\begin{equation}
    S_2(\tau) \sim (\varepsilon \tau U )^{2/3}.
\end{equation}
If sweeping is not negligible even in the Lagrangian frame, this scaling can be expected to hold for time scales $\tau \lesssim \tau_L$, while for $\tau \ll \tau_L$ we can expect to recover the predicted Lagrangian scaling $S_2(\tau) \sim \varepsilon\, \tau$. Note that this behavior is also compatible with a scaling of the Lagrangian frequency spectrum $E(f)\propto f^{-5/3}$ at intermediate frequencies, and of $E(f)\propto f^{-2}$ only for sufficiently large frequencies as seen in Fig.~\ref{lag:energy_spec}.

In Fig.~\ref{lag:S2_tau23} we show the second order Lagrangian structure functions $S_2(\tau)$ compensated by this prediction, both for all experimental datasets, as well as comparing EXP50 data with the DNS. A range of time lags with approximately constant compensated structure functions can be seen; its width being approximately independent of the Reynolds number. This is consistent with the interpretation that the observed scaling of $S_2(\tau)$ is associated with large-scale flow effects, as the structure of the mean flow is fixed by the geometry of the setup in the experiments or by the volumetric forcing in the DNS. Indeed, energy is injected at approximately the same scales in all experimental runs, and the mean flow has little dependence with the Reynolds number when turbulence has reached a fully developed state, as already discussed. The observed plateau is slightly wider for the horizontal velocity component, which is also consistent with the anisotropy generated by the mean flow at the largest scales (large-scale vortices in von K\'arm\'an and Taylor-Green flows display larger correlation length and correlation time in the horizontal than in the vertical direction). Both the experiments and the DNS show similar levels of anisotropy (although, as before, the DNS results are slightly more anisotropic).  
All these observations reinforce the idea that advection of the tracers by the mean Eulerian flow affect second order Lagrangian statistical measurements, and could also be the source of the very limited $S_2(\tau) \sim \varepsilon\, \tau$ scaling reported in previous experiments \citep{ouellette2006, biferale2008}.

\begin{figure}
  {\includegraphics[width=1\textwidth]{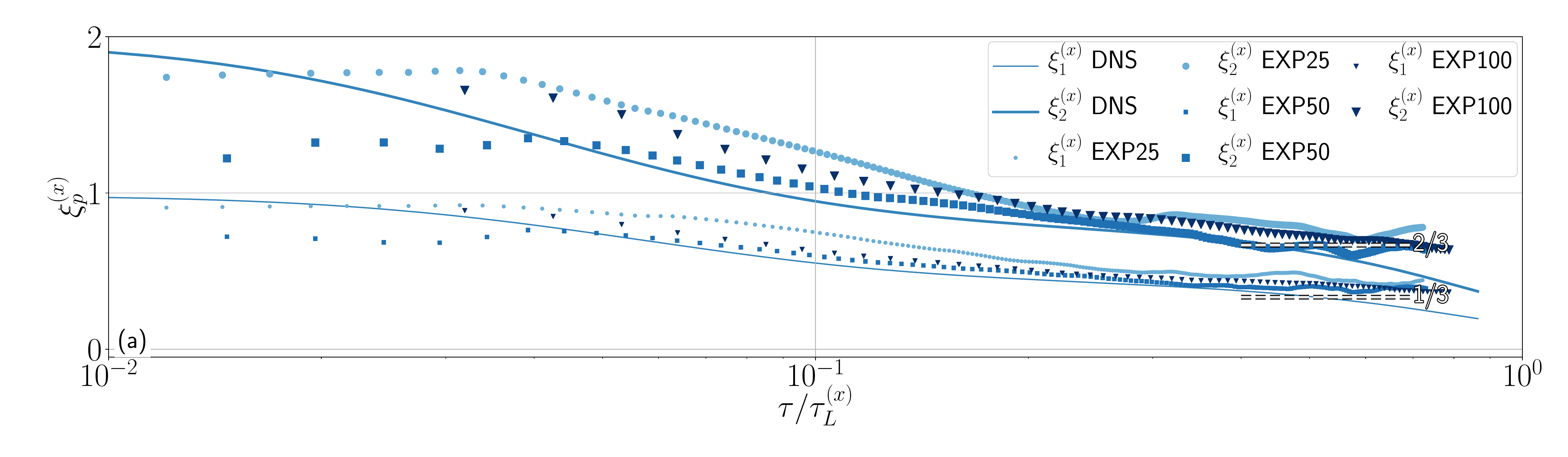}}
  \hfill%
  {\includegraphics[width=1\textwidth]{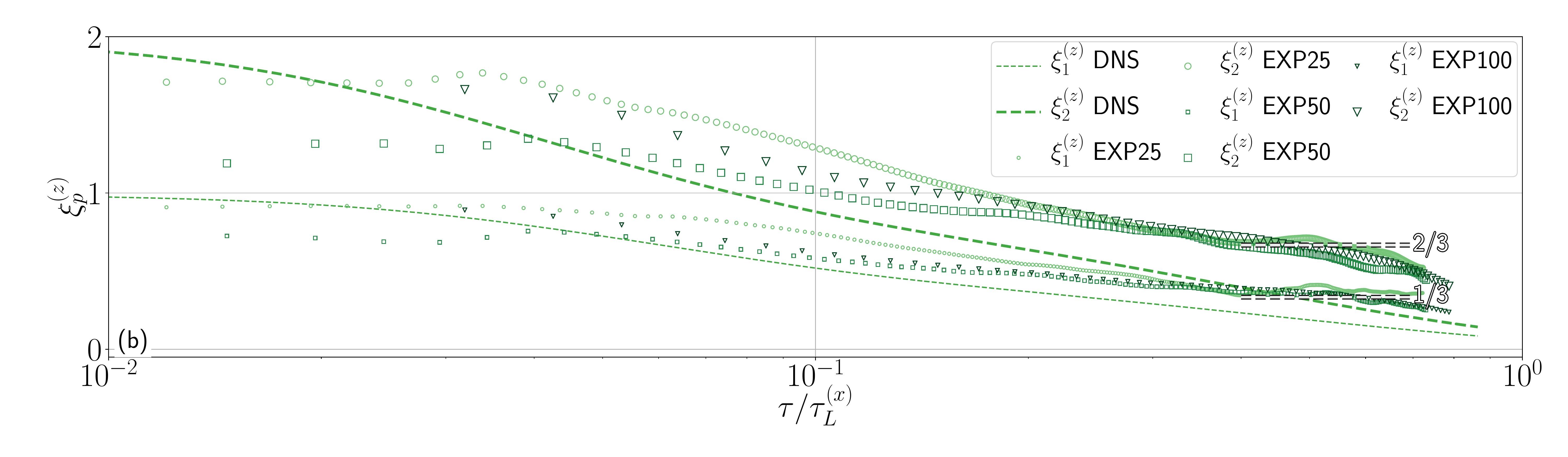}}
  \caption{Logarithmic derivatives of the particle velocity structure functions of order $p=1$ and $p=2$ (see labels in the inset). Each individual panel gathers both experimental and numerical results for a given Cartesian component of the velocity; $x$ component in panel (a), and $z$ component in panel (b). Values of $1/3$ and $2/3$ are indicated as references by the horizontal dashed lines.}
  \label{lag:exponents}
\end{figure}

The anomalous behavior of the second order Lagrangian structure function, and the impact of the mean flow, can be also observed in other statistical moments of the tracers' velocity. Statistics of the tracers' for other orders can be accessed, e.g., considering the Lagrangian structure function of order $p$,
\begin{equation}
    S_p^{(i)} (\tau) = \langle |v_i (t+\tau) - v_i(t)|^p \rangle.
\end{equation}
In order to asses the statistical properties locally, we compute 
the local scaling exponent (LSE) given by the logarithmic derivative of the corresponding structure function: 
\begin{equation}
    \xi_p^{(i)}(\tau) = \frac{d\  \log \: S_p^{(i)}(\tau) }{d\ \textrm{log}(\tau)},
\end{equation}
for $p = 1$ and $2$. LSEs are more conducive to analysing scaling properties scale-by-scale as they are expected to remove large-order non-universal contributions coming from the overall prefactors in the structure functions (see, e.g., \citep{Benzi2010}). 
In particular, the case $p=2$ provides us with an alternative way to analyze the behavior of $S_2^{(i)}(\tau)$. The results for all the experiments and the simulation are shown in Fig.~\ref{lag:exponents}, each panel corresponding to a different Cartesian velocity component. In the Lagrangian inertial range and in the absence of intermittency, $\xi_p \approx p/2$ is expected, whereas in the sweeping-dominated range we anticipate $\xi_p \approx p/3$. For $\tau/\tau_L^{(x)} \approx 0.1$ the local slopes cross the Lagrangian inertial range prediction, while for $\tau/\tau_L^{(x)} \lesssim 1$ the local slopes are consistent with sweeping for both $p=1$ and $2$. Moreover, and as noted for other quantities, a better scaling is seen (for both ranges) in the horizontal component of the velocity, both for the experiments and the DNS, with clear differences between the two velocity components due to anisotropy. 

\subsection{Statistics of velocity fluctuations in subregions \label{sec:lag_subregions}}

\begin{figure}
  {\includegraphics[width=0.5\textwidth]{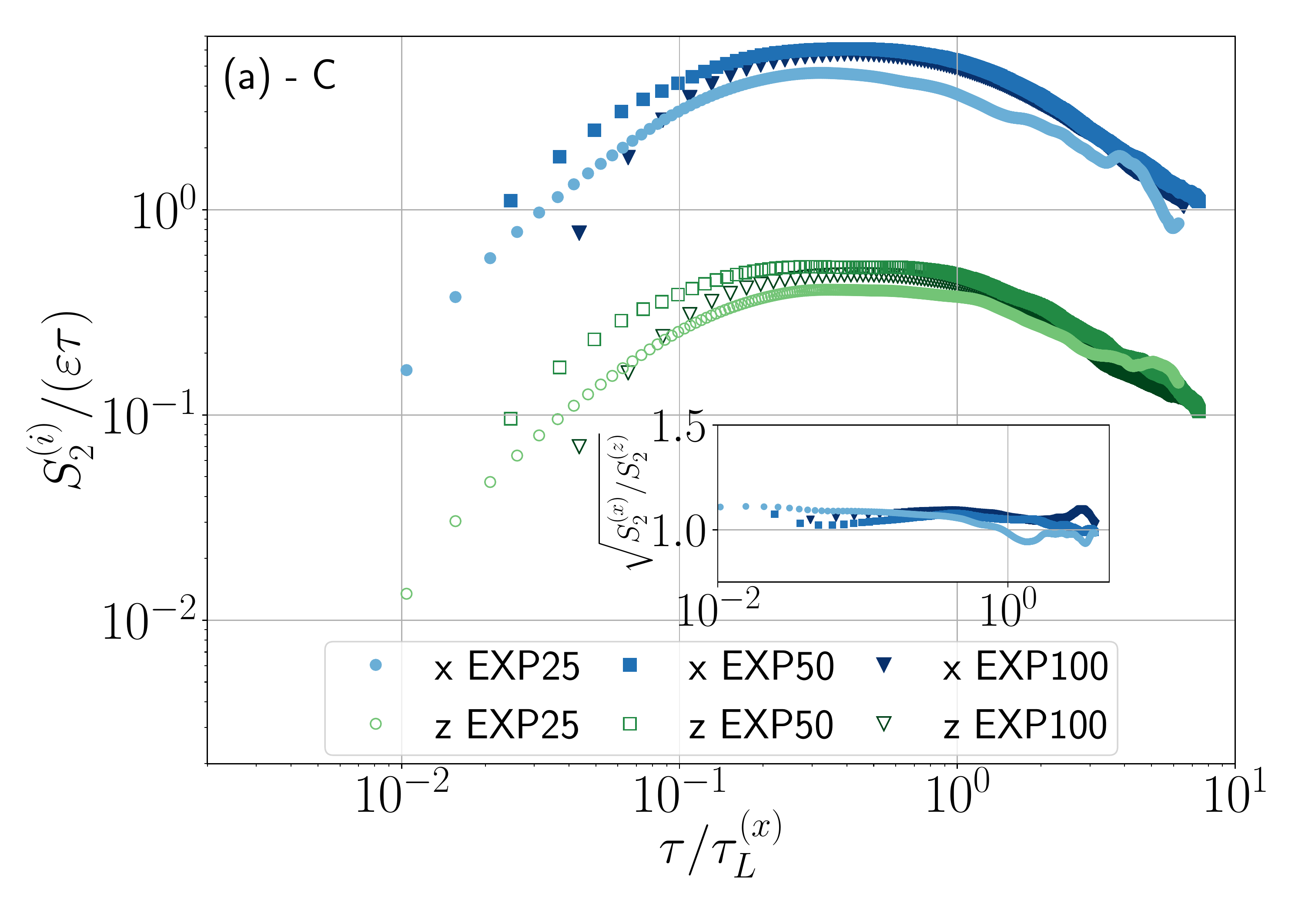}}%
  \hfill%
  {\includegraphics[width=0.5\textwidth]{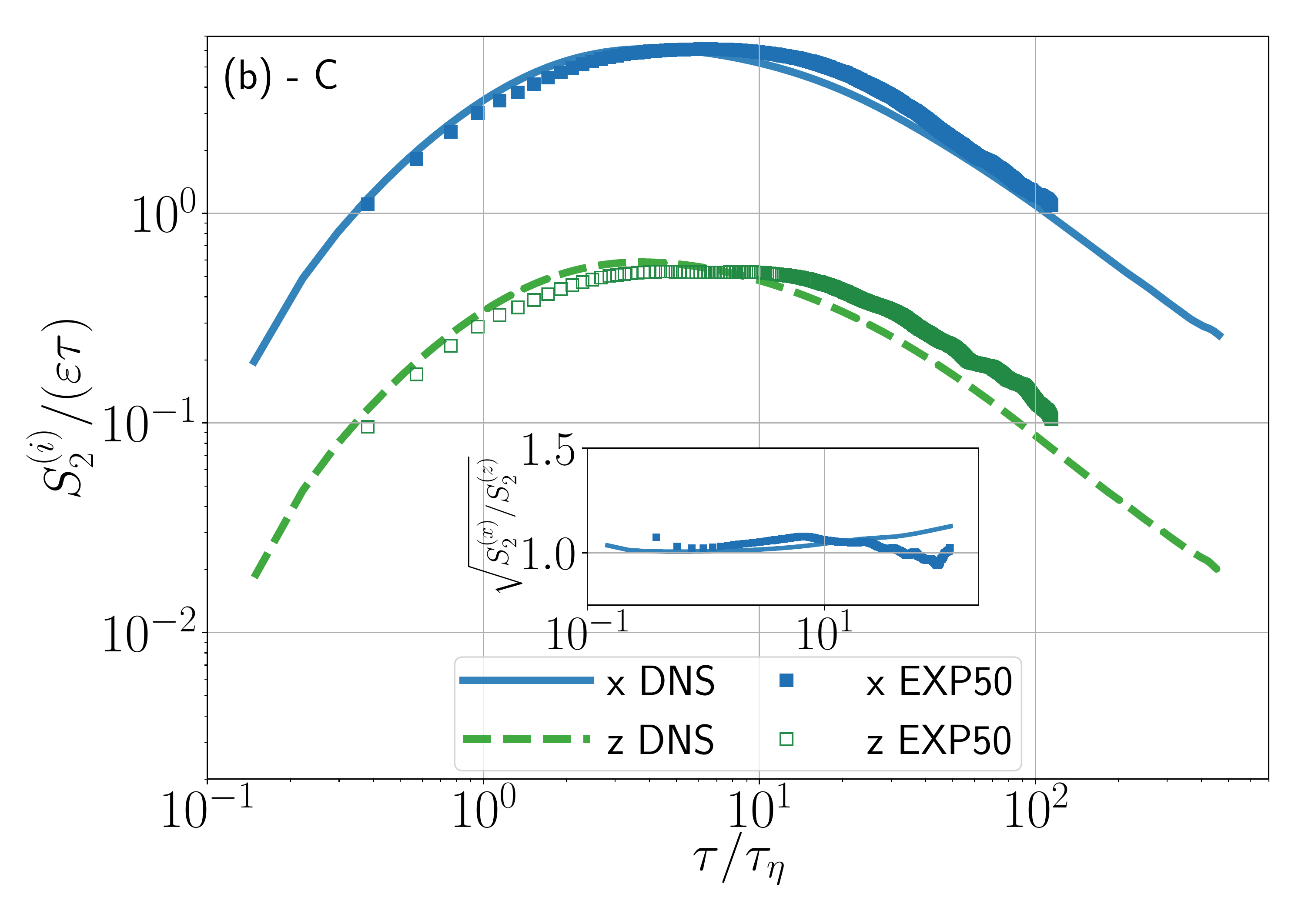}}
  \caption{Second order tracers' structure functions for each Cartesian velocity component, conditioned to particles located in the central subregion of the cell (subregion C), and compensated by the Lagrangian inertial range prediction $\varepsilon\tau$. Panel (a) shows experimental data, and panel (b) shows DNS data compared with EXP50. In both figures the $z$ component is displayed with an arbitrary vertical displacement of $10^{-1}$. The insets show $\sqrt(S_2^{(x)}/S_2^{(z)})$ using linear scale in the vertical axis.}
  \label{lag:S2_C}
\end{figure}

In light of the presence of two different behaviors in the second order Lagrangian structure functions (one compatible with $S_2 \propto \tau$ and the other with $S_2 \propto \tau^{2/3}$), and considering that the sweeping by the large scale flow may be the cause of the former behavior, we compute second order velocity structure functions for the tracers but now restricting the tracers' positions to two different subregions of the cell. The choice of the two subregions attempts to clarify the effect of the mean flow. In the experiments, one subregion comprises the central horizontal $1/3$ of the observed volume, and includes (on the average) the shear layer in which the von K\'arm\'an flow can be expected to be less inhomogeneous and anisotropic. This subregion will be labeled ``C'' in the following, for ``central'' region. The other subregion includes the top $1/3$ and bottom $1/3$ of the volume closer to the disks. In these regions the effect of the mean flow can be expected to be more important as they are closer to the forcing mechanism. This whole subregion well be labeled as ``D'' as it comprises the fluid close to the disks. In the DNS, following the same procedure, we separate each Taylor-Green cell in two subvolumes, $1/3$ corresponding to the central part including the shear layer (and where the external forcing is minimal), and the top $1/3$ and bottom $1/3$ (where the Taylor-Green forcing is maximal). For these subregions we use the same labels as in the experiments.

\begin{figure}
  {\includegraphics[width=0.5\textwidth]{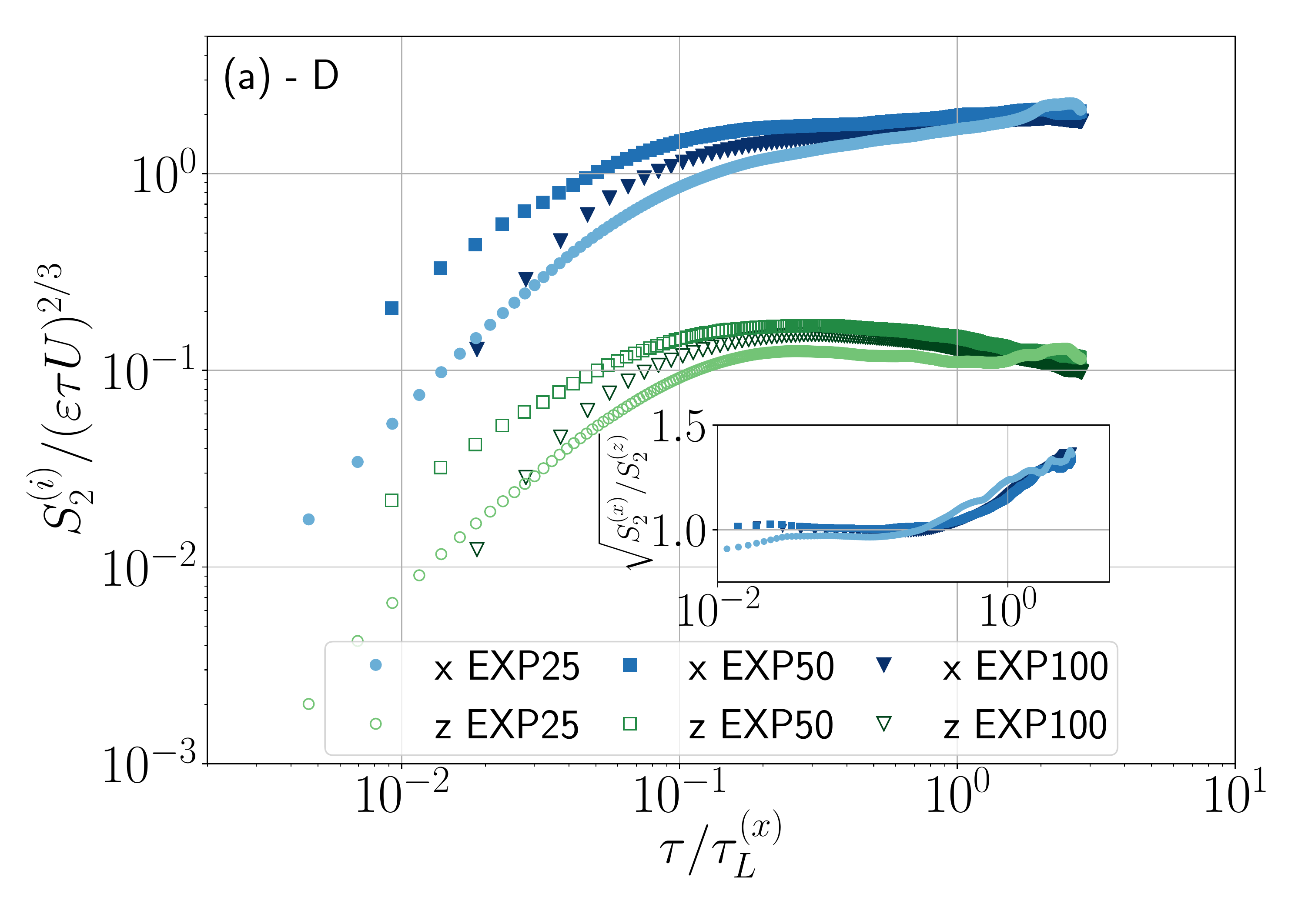}}%
  \hfill%
  {\includegraphics[width=0.5\textwidth]{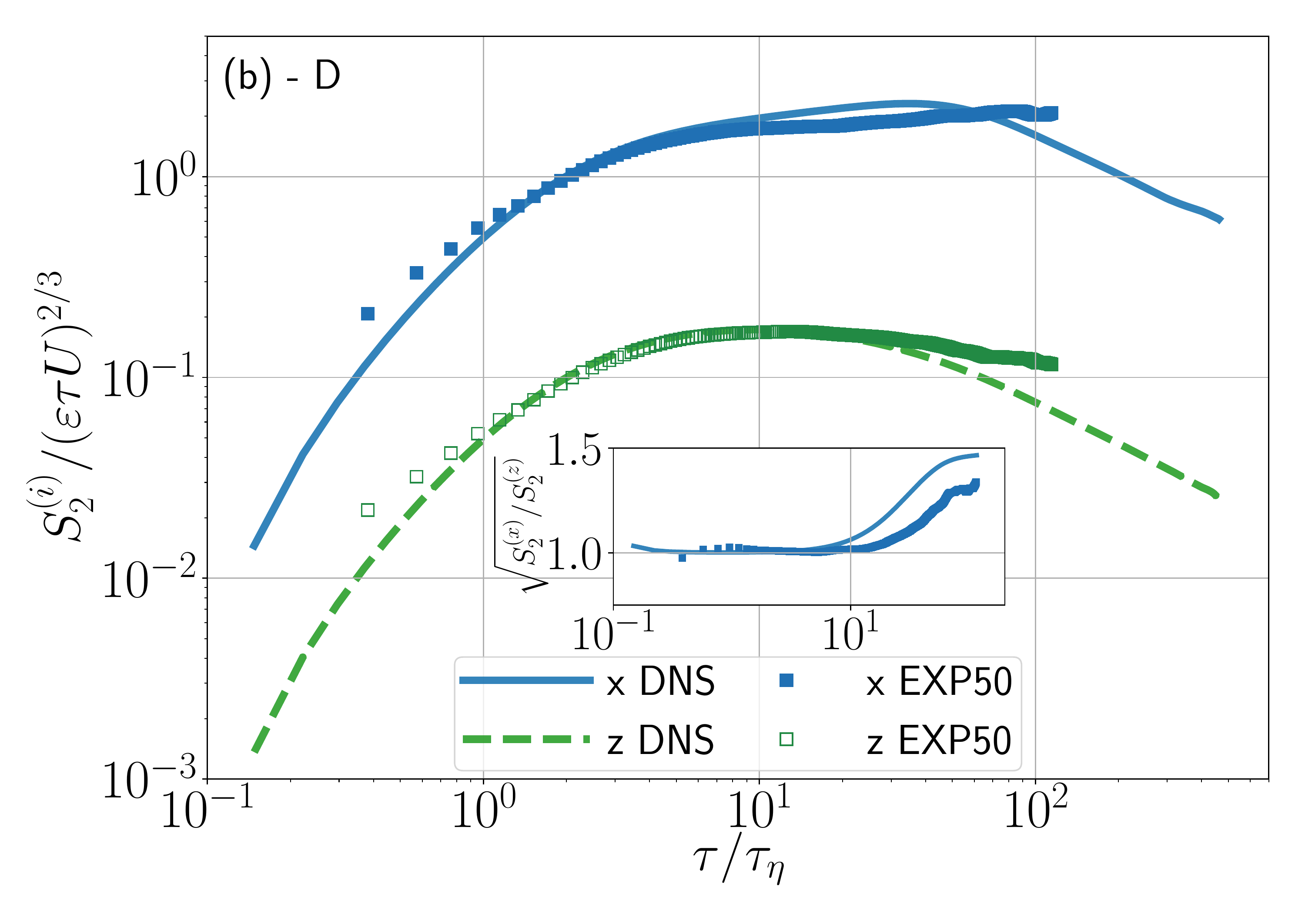}}
  \caption{Second order tracers' structure functions for each Cartesian velocity component, conditioned to particles located in subregion D, and compensated by the sweeping-dominated prediction $(\varepsilon\tau U)^{2/3}$. Panel (a) shows experimental data, and panel (b) shows DNS data compared with EXP50. In both figures the $z$ component is displayed with an arbitrary vertical displacement of $10^{-1}$. The insets show $\sqrt(S_2^{(x)}/S_2^{(z)})$ using linear scale in the vertical axis.}
  \label{lag:S2_D}
\end{figure}

We first consider $S_2(\tau)$ in subregion C. Figure \ref{lag:S2_C}(a) shows the experimental results compensated by $\varepsilon\tau$, the expected behavior in the Lagrangian inertial range of homogeneous and isotropic turbulence. Even though the structure functions display more fluctuations than when computed for all trajectories (as a result of having less statistics), a broader plateau than in Fig.~\ref{lag:S2_tau} can be seen, and for larger time lags $\tau/\tau_L^{(x)}$ (note also that the values of $\tau_L^{(x)}$ used to normalize times in this figure were recomputed for the trajectories in this subregion, and are approximately half the value of the autocorrelation time in the entire domain). The width of the plateau seems also independent of whether $S_2^{(x)}(\tau)$ or $S_2^{(z)}(\tau)$ are considered. As shown by the ratio $\sqrt (S_2^{(x)}/S_2^{(z)})$, subregion C seems also more isotropic than the flow in the entire volume. This is also confirmed by the ratio of the autocorrelation times $\tau_L^{(x)}/\tau_L^{(z)}$ computed only for trajectories in subregion C in the three experiments: $\tau_L^{(x)}/\tau_L^{(z)} = 1.02$ in EXP25, $0.94$ in EXP50, and $1.07$ in EXP100 (compare these values with those listed for the entire domain in Table \ref{Lag_table}). Figure \ref{lag:S2_C}(b) then compares $S_2/(\varepsilon\tau)$ in the same subregion for the DNS and for EXP50. A similar behavior is observed: a good collapse of numerical and experimental data, a plateau, and a similar reduction in the flow anisotropy (see the inset in this figure). Finally, in this subregion, when structure functions are compensated by $(\varepsilon \tau U)^{2/3}$ (not shown), the sweeping-dominated scaling range is significantly shortened when compared with the data for the whole volume in Fig.~\ref{lag:S2_tau23}.

Figure \ref{lag:S2_D} show $S_2(\tau)$ compensated by $(\varepsilon \tau U)^{2/3}$ for the experiments and for the DNS, but now conditioned to particles in subregion D. For all the datasets, a plateau is observed for both Cartesian velocity components, and for the $x$ component of the velocity in a broader range of time scales than that observed in Fig.~\ref{lag:S2_tau} for the entire volume. Note also that the ratio $\sqrt (S_2^{(x)}/S_2^{(z)})$ indicates the flow is more anisotropic in this subregion. Again, a good agreement is seen between EXP50 and the DNS data, with a broader plateau in the structure function of the $x$ velocity component, and with the DNS data being more anisotropic for the largest time lags than the experimental data. And as in the previous case, the data in this subregion shows no clear scaling range when compensated by $\varepsilon \tau$ (not shown).

These results give further evidence that the origin of the $\sim (\varepsilon \tau U)^{2/3}$ scaling is associated with the effect of the mean flow, as the scaling is stonger in subregions in which the mean flow is stronger. This also implies that, once the effect of the mean flow has been mitigated (e.g, by condering regions in the cell far from the injection mechanism, or by carefully removing it as done in Refs.~\citep{machicoane2016, huck2019}), a more clear $\sim \varepsilon \tau$ scaling could be identified.

\subsection{Acceleration spectra}

As discussed in the previous sections, the reason for the anomalous behavior of the energy spectra $E(f)$ and of the structure functions $S_2(\tau)$ observed in the literature and in our datasets seems to be the contamination of the scaling by mean flow effects. It has already been proposed by other authors (albeit not identifying the source of contamination to sweeping) that second order structure functions mix low frequency fluctuations with inertial range fluctuations \citep{huang2013}, and that once the large scale contamination is removed, a plateau should emerge even at $Re_\lambda \approx 400$ \citep{lanotte2013}. As a result, computation of the so-called acceleration spectrum was proposed as another solution, as this spectrum is expected to better disentangle the contributions from different time scales, and as a result show a clearer Lagrantian inertial range scaling \citep{sawford2011}. Recently, the acceleration spectrum was computed for data from von K\'arm\'an flows \citep{huck2019}, and it was found that the anisotropy of the flow is indeed contained in low frequencies, and that the spectra are isotropic for frequencies in the dissipative range. 

\begin{figure}[b]
  {\includegraphics[width=0.5\textwidth]{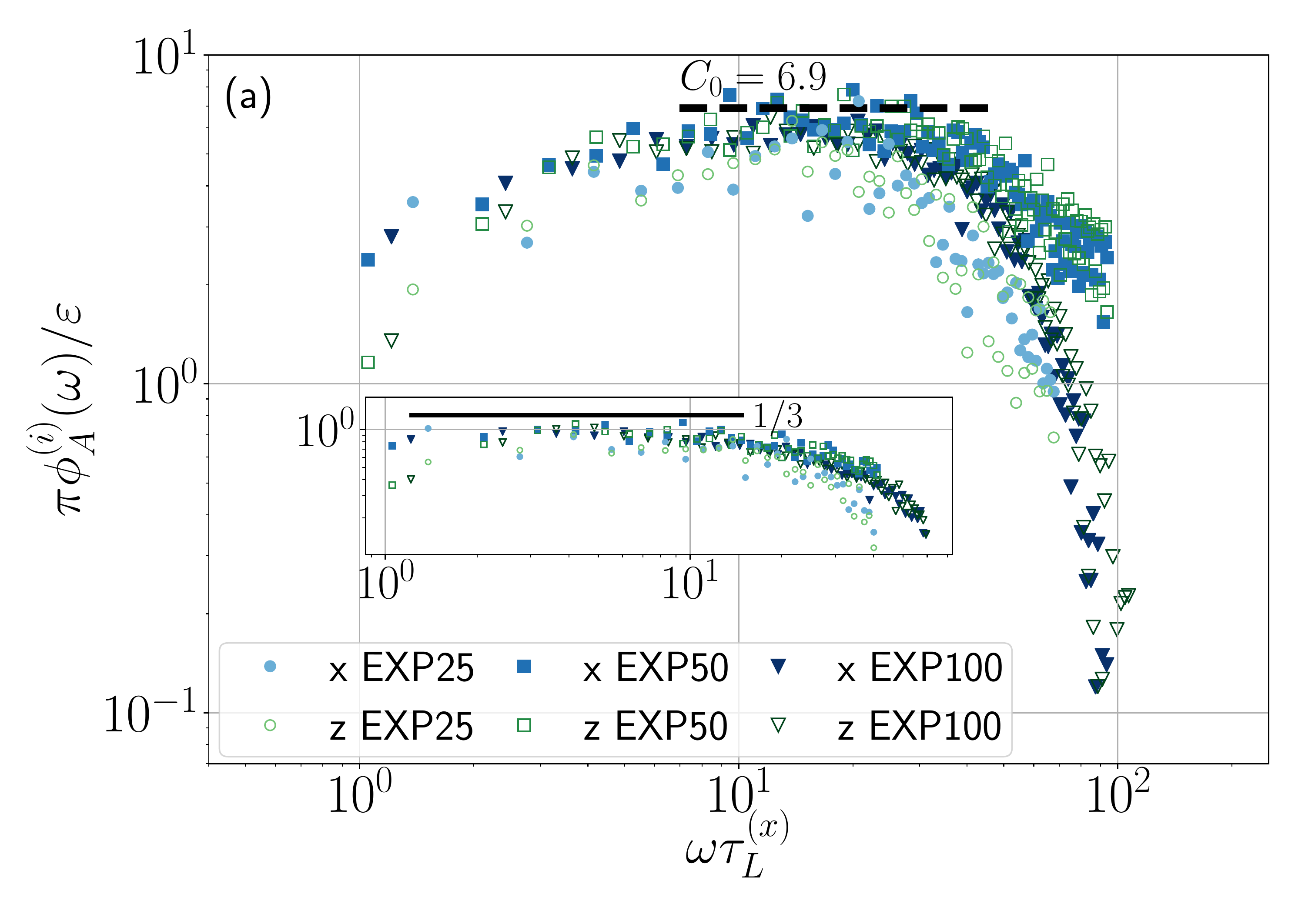}}%
  \hfill%
  {\includegraphics[width=0.5\textwidth]{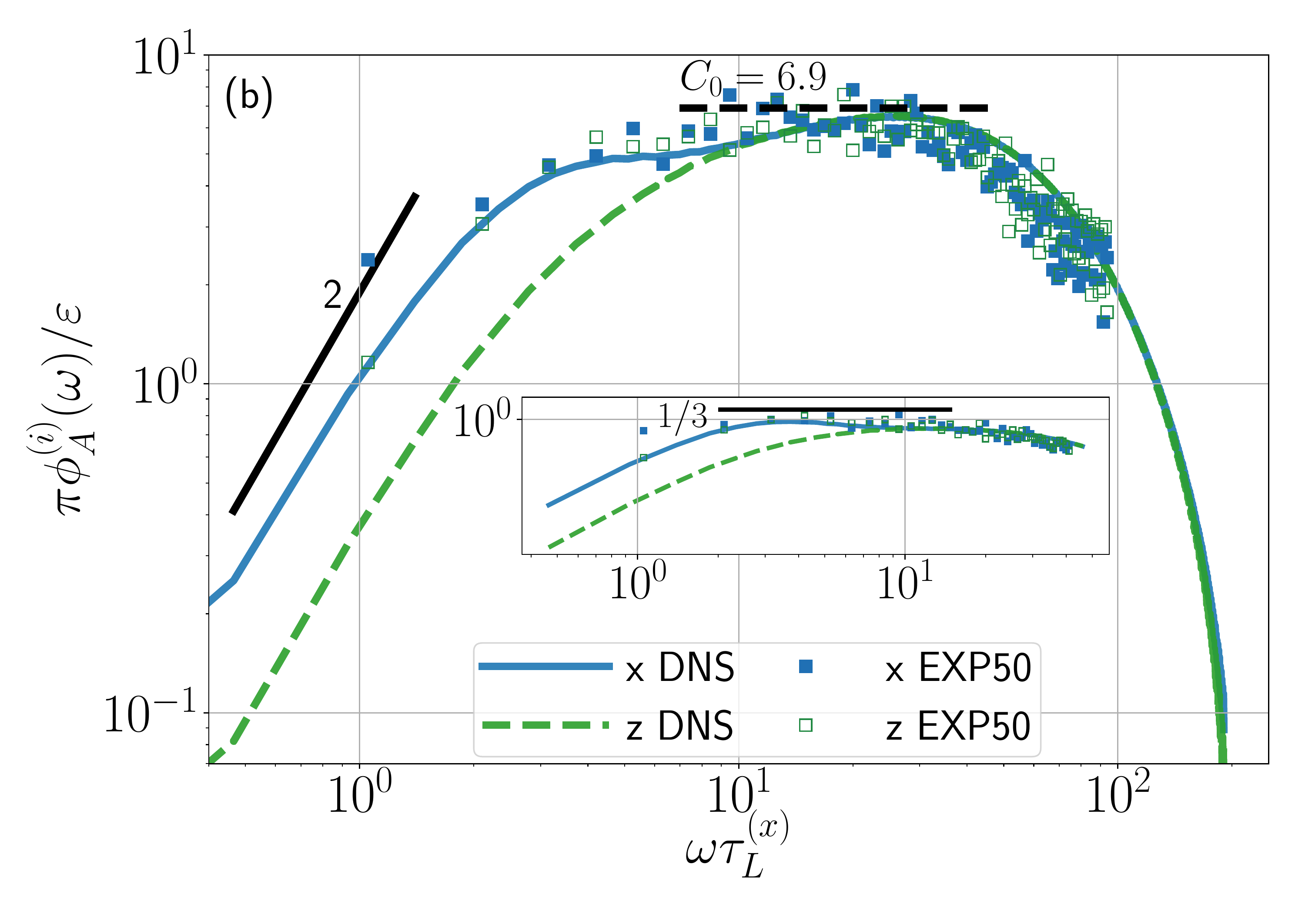}}
  \caption{Particle acceleration spectra normalized by the energy injection rate $\varepsilon$, in (a) experiments, and (b) EXP50 and the DNS. The dashed horizontal line indicates the amplitude $C_0 = 6.9$, which corresponds to the asymptotic value expected for the amplitude of the acceleration spectrum in the Lagrangian inertial range. In panel (b) a $\omega^2$ slope is also indicated as a reference. The insets in both panels show the same spectra compensated by a power law with exponent $1/3$, corresponding to sweeping.}
  \label{lag:acc_spec}
\end{figure}

The tracers' one-dimensional acceleration spectrum, computed from the particles' acceleration autocorrelation function $C_A^{(i)}(\tau)$, is defined as
\begin{equation}
    \phi_A^{(i)}(\omega) = \frac{2}{\pi} \int\limits_0^\infty  C_A^{(i)}(\tau)\, \cos(\omega \tau) d\tau =\frac{2}{\pi} \omega^2 \int\limits_0^\infty  C_v^{(i)}(\tau)\, \cos(\omega \tau) d\tau , \label{eq:acc_spec}
\end{equation}
where $\omega = 2\pi f$ as before. Note the second expression allows computation of $\phi_A^{(i)}(\omega)$ directly from the velocity autocorrelation function $C_v^{(i)}(\tau)$. This spectrum is expected to scale as $\pi \phi_A(\omega)/\varepsilon \propto \omega^2$ for $\omega \to 0$, and as $\pi \phi_A(\omega)/\varepsilon = C_0$ in the inertial range, i.e., for $\tau_L^{-1} \ll \omega \ll \tau_\eta^{-1}$. The value of $C_0$, estimated by extrapolating the peak in $S_2(\tau)/\varepsilon\tau$, is found to be $C_0 \approx 6.9$ (see \citep{sawford2011}).
As mentioned before, in previous studies it was noted that this spectrum converges much faster to a plateau than $S_2(\tau)$, and thus it should be a better indicator of the existence of a Lagrangian ``Kolmogorov-like" inertial range.

The normalized acceleration spectra associated to our experimental datasets are shown in Fig.~\ref{lag:acc_spec}(a), while the DNS data is compared with results from EXP50 in Fig.~\ref{lag:acc_spec}(b). In all cases, spectra are computed from particle trajectories in the entire volume. For both the experiments and the simulation, we obtain a plateau spanning almost one decade in frequency range, with an amplitude close to the predicted value for $C_0$. The extension of the plateau appears to grow with the Reynolds number. For low frequencies, a range compatible with a power law with exponent $1/3$ is also visible in all the datasets (see the insets with compensated spectra in Fig.~\ref{lag:acc_spec}). The presence of this power law is consistent with the $-5/3$ range visible in the velocity power spectra, and the sweeping-like scaling in the Lagrangian structure functions. For the DNS, where larger times have been sampled, at even smaller frequencies ($\omega\tau_L^{(x)} \approx 1$) the spectrum grows as $\omega^2$, as expected. More importantly, the numerical and experimental data show a remarkable collapse, sharing characteristics such as the growth at low frequencies, a comparable span of time scales for the inertial range, and a similar drop at high frequencies. The large-scale anisotropy of the flow can be also identified in these curves: the plateau is wider for the horizontal component although, as in all previous cases, the anisotropy at large scales is more pronounced in our DNS than in the experiments.

\section{Inertial particles}

The good agreement between the statistical properties of tracers' velocity and acceleration in the von K\'arm\'an experiments and in the Taylor-Green simulations, even when considering large scales associated to the mean flow and in spite of the conspicuous differences in the boundary conditions, encourages us to perform comparisons between inertial particles in the laboratory flow and the simulations. The motivation behind the comparisons in this section is to explore the possibility of performing validation of effective models for large particles (and their parameters) using statistical comparisons between two reminiscent flows. Thus, we consider a large particle in the experiment, and we compare it with point particles in simulations that evolve according to Eq.~(\ref{eq:iner}), looking for effective values of the Stokes time that make the statistical properties between the two cases comparable. Other models for the particles could be considered, as, e.g., the Maxey-Riley equation with all corrections up to first order in the particles' radius as given by Eq.~(\ref{eq:MR}), or to order $R^2$. However, even in the latter case the equations hold only for particles with $Re_\textrm{p}\ll 1$, while in many cases (including our experiment) $Re_\textrm{p} \gtrsim 1$. As a result, instead of considering other effects, we will regard Eq.~(\ref{eq:iner}) as an empirical model with one effective coefficient, assuming the velocity of large particles tries to match the fluid velocity with an effective relaxation time $\tau_p$. If this can be succesfully done, in future studies we will consider other models with more effective parameters to take into account, e.g., the effect of gravity and of an effective added mass \cite{maxey1983}.

Comparisons of inertial particles in DNSs and von K\'arm\'an flows taking into account some of these effects have already been reported in the bibliography (see, e.g., \citep{Volk2008}), where simulations of isotropic and homogeneous turbulence showed good agreement with the experimental data. The authors also pointed out effects induced by the finite size of the particles in their comparisons, that were not captured by the simulations. In our case, we will focus instead on exploiting the Eulerian and Lagrangian similarities of the Taylor-Green and von K\'arm\'an flows, to test which effects of the experimental particle's dynamics are well captured by the simulations even when using a simpler model, and to focus mostly on the effect of the large scale flow in the statistics of the particles in both the numerics and the experiments.

To this end, on the one hand the dynamics of a large particle (with diameter $6$ mm) is studied in the experiment, using $f'_0 = 50$ rpm. As shown in the previous section, this rotation frequency generates an experimental flow with a $Re_\textrm{part}$ value similar to that reached in the simulations. On the other hand, in the simulations, four values of $\tau_p$ are explored; namely: $\tau_p = 0.2,\, 0.5,\, 1.5$, and $3.0$ (in dimensionless units). For ease of reference, simulations for each of these values of $\tau_p$ will be termed DNS0.2, DNS0.5, DNS1.5, and DNS3.0, respectively. Inertial particles were evolved in the same turbulent flow employed in the study of Lagrangian tracers described in Section~\ref{sect:lag-tracers}. 

\begin{table}
\begin{ruledtabular}
\begin{tabular}{l c c c cc c c c c c c}
\hfill
    Dataset & $f'_0$ & $\tau_p^*$ & $R/\eta$& $Re_\textrm{p}$ &$St_\eta$ & $St_\textrm{int}$ & $St_R$ & $\tau_L^{(x)} f_0$ & $\tau_L^{(x)}/\tau_L^{(z)}$ & $T^{(x)} f_0$ & $T^{(x)}/T^{(z)}$ \\
    \hline
            & [rpm] & [ms] &  &  \\
    EXP6       & 50        &  $100$  & $29.5$ & $3.5$	&$274$&  $1.24$  & $9.5$  & $0.38$ & $1.19$  & $0.18$ & $1.26$\\
    \hline
    DNS0.2     &  $1/2\pi$ & 0.2   & $10.2$   & $4.4$ &$4.7$ &  $0.029$  & $4.7$ & $0.43$ & $1.29$  & $0.16$ & $1.39$\\       
    DNS0.5     &  $1/2\pi$ & 0.5   & $37.1$   & $16.0$&$11.1$&  $0.069$  & $11.1$ & $0.48$ & $1.22$ & $0.19$ & $1.26$ \\       
    DNS1.5     &  $1/2\pi$ & 1.5   & $201.0$  & $27.7$&$34.3$&  $0.22$  & $34.3$ & $0.58$ & $1.09$  & $0.27$ & $1.15$\\       
    DNS3.0     &  $1/2\pi$ & 3.0   & $574.1$  & $79.0$&$69.1$&  $0.43$  & $69.1$ & $0.72$ & $1.00$  & $0.36$ & $1.09$\\  
\end{tabular}
\end{ruledtabular}
\caption{Parameters of experiments and simulations with inertial particles. The DNS values are dimensionless. The flow is characterized by the Lagrangian/tracers measurements (see Table~\ref{Lag_table}). For the experiment, $f'_0=60 f_0$, where $f_0$  corresponds to the frequency of the disks measured in Hz, whereas for the DNS, it corresponds to the frequency associated to a large-scale eddy turn over time. Then inertial particles' response time is denoted by $\tau_p^*$, and corresponds to the time that best describes the particles' dynamics when compared with the simulations (see text). In the experiments, $\tau_p^* = \tau_R = R^{2/3}/ \varepsilon^{1/3}$, $R$ being the particle's radius; in the DNSs $\tau_p^*=\tau_p$. Equivalently, the effective radius of the particles in the simulations is defined as $R = \tau_p^{3/2} \varepsilon^{1/2}$. $Re_\textrm{p} = R\lvert{\bf u} - {\bf v}\lvert / \nu$, where ${\bf u}$ and ${\bf v}$ are the flow and particle velocities, respectively. $St_\eta = T_p/\tau_\eta$, where in the experiments $T_p = (2/9\nu) R^2 (\rho_p/\rho + 1/2)$, and in the DNSs $T_p = \tau_p$. $St_\textrm{int} = T_p/\tau_\textrm{int}$, where $\tau_\textrm{int} = L/U$. $St_R$ represents the Stokes number based on $\tau_R$, and is given by $St_R = \tau_R/\tau_\eta$ ($=\tau_p/\tau_\eta$ in the DNSs). $\tau_L^{(x)}/\tau_L^{(z)}$ is the ratio of the correlation times $\tau_L^{(i)}$, obtained from the zero-crossing of the normalized autocorrelation functions $R_L^{(i)}(\tau)$. $T^{(x)} f_0$ and $T^{(x)}/T^{(z)}$ are dimensionless measurements of the ``particle integral time" $T^{(i)} = \int_0^{\tau_{95}} R_L^{(i)}(\tau) \: d\tau$, with $R_L^{(i)}(\tau_{95}) = 0.05$.}
\label{Iner_table}
\end{table}

The different values of the Stokes number for the DNSs and the experiment, calculated from Eqs.~\eqref{eq:St} and \eqref{eq:StR} are shown in Table~\ref{Iner_table}. As is usually the case in studies of particle-laden turbulent flows, the values of the Stokes numbers in the experiment and in the simulations can be vastly different depending on the definition used. The particle Stokes time $T_p$ computed as in Eq.~(\ref{eq:St}) yields $T_p=2840$ ms, resulting in Stokes numbers $St_\eta$ and $St_\textrm{int}$ with differences of several orders of magnitude between the experiment and the DNS data (even though, as will be shown next, statistical results between the experiment and the simulations are compatible in many cases). Instead, the Stokes number $St_R$ defined as in Eqs.~\eqref{eq:StR} and \eqref{eq:tau_R} (based on $\tau_R=100$ ms for the experiments) results in values which are comparable with those of the simulations showing closest agreement to the experimental data. Using $St_R$, values of the experimental data lay between DNS0.2 and DNS0.5, the numerical simulations with $\tau_p = 0.2$ and $0.5$ respectively. Note that the ratio $R/\eta$, and the value of $Re_\textrm{p}$ for each dataset, as well as other dimensionless numbers in Table~\ref{Iner_table}, are also of the same order of magnitude for these datasets. Indeed, from $R$ we can estimate $Re_\textrm{p}= 3.5$ for the experiment. Keeping in mind that $\tau_p$ and all Stokes numbers for the DNSs are effective quantities (as we are considering a particle with $Re_\textrm{p} \gtrsim 1$), we can also estimate from the effective value of $R$ in the simulations the effective value for $Re_\textrm{p}$. Based on this number the experimental data seem to be close to DNS0.2.

\subsection{Velocity probability density functions}

We begin by studying the probability density functions (PDFs) of the Cartesian components of the velocity of the inertial particles. Even though a Gaussian distribution is expected in the case of tracers, deviations resulting, e.g., in sub-Gaussian statistics have been reported in the case of inertial/finite-sized particles in experiments \citep{volk2011}, and in numerical simulations of homogeneous and isotropic turbulence \citep{homann2010}. 

\begin{figure}
  {\includegraphics[width=0.5\textwidth]{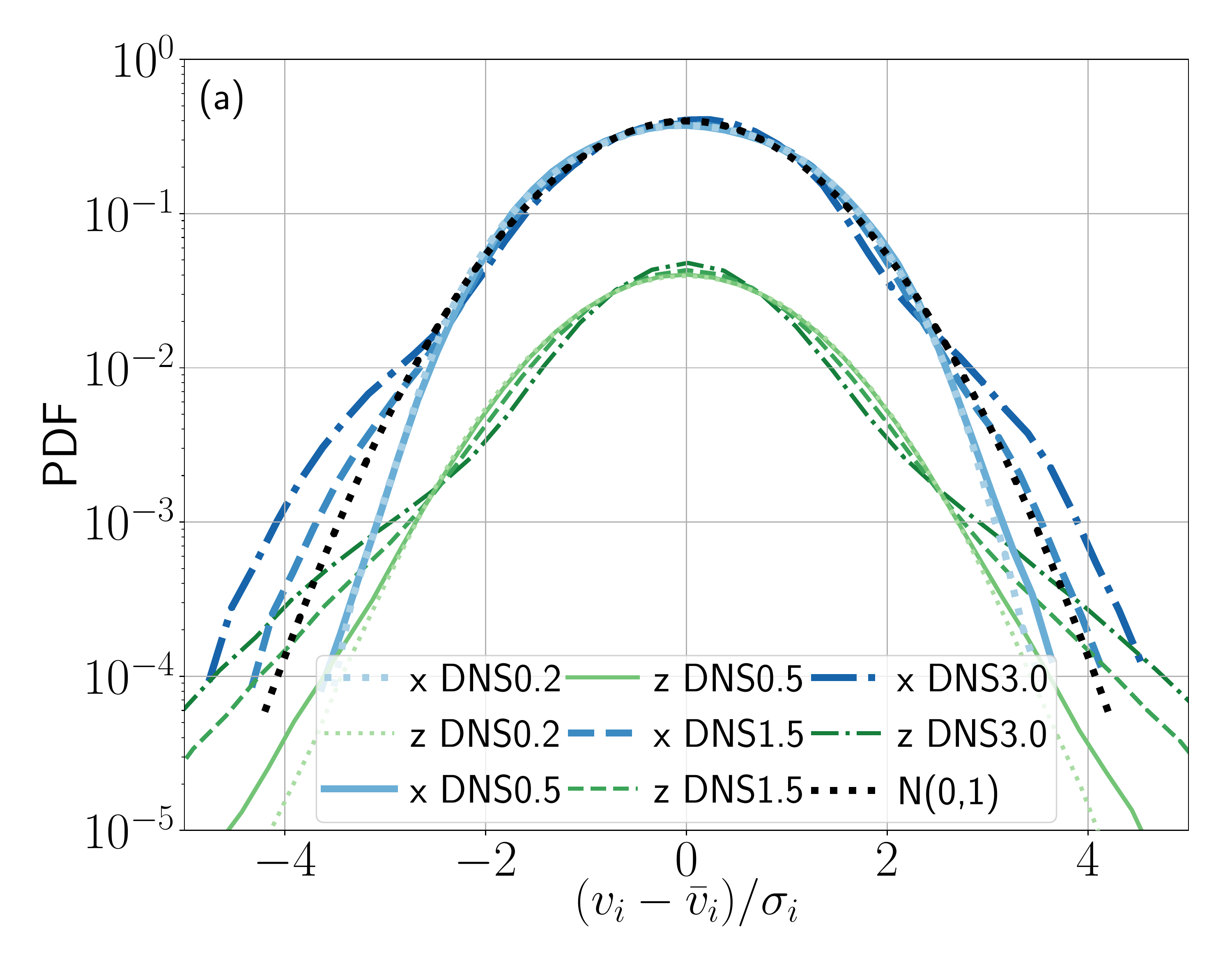}}%
  \hfill%
  {\includegraphics[width=0.5\textwidth]{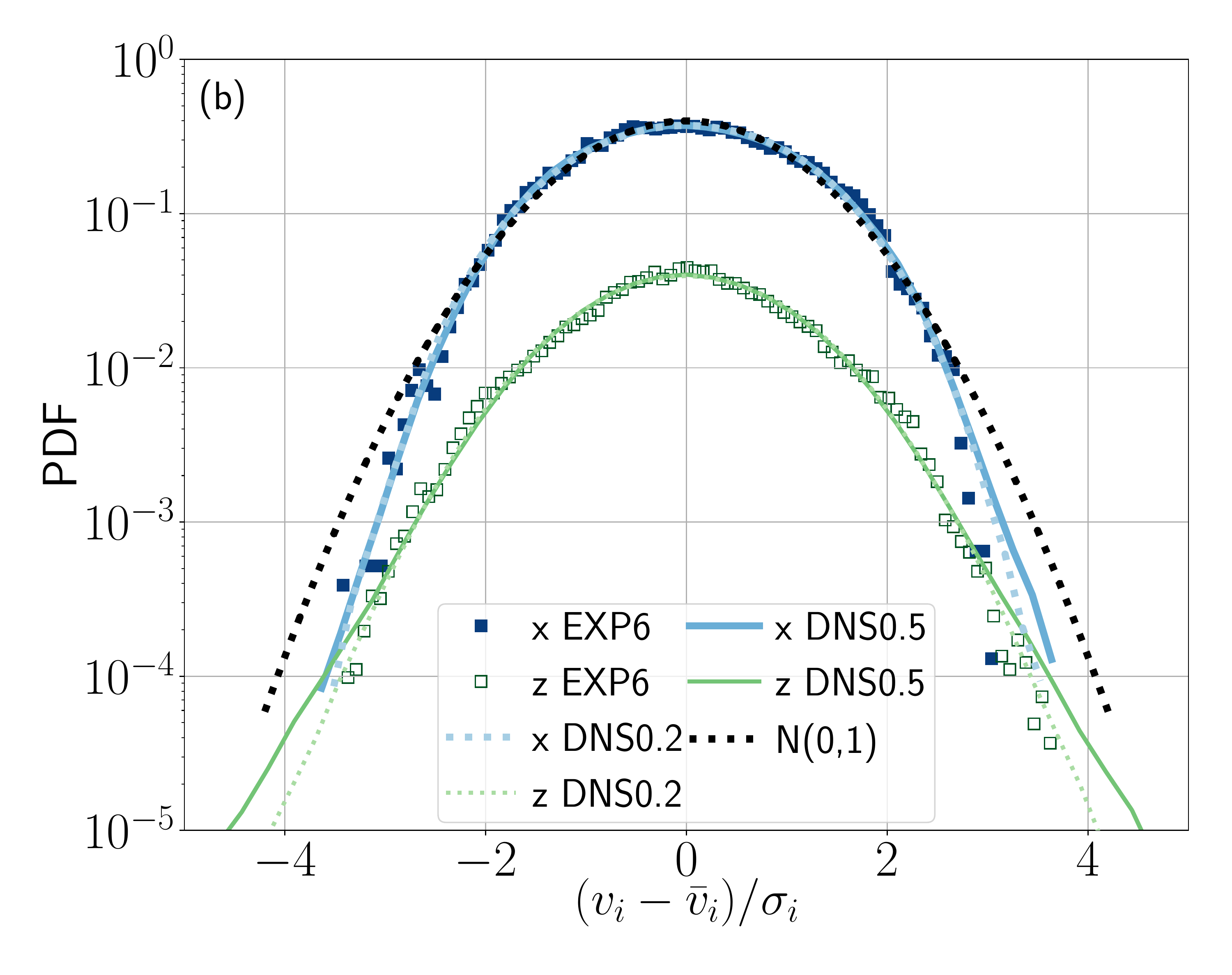}}
  \caption{Normalized probability distribution function of the Cartesian components of the inertial particles velocity. Panel (a) gathers all four DNSs, while panel (b) compares results for EXP6, DNS0.2, and DNS0.5. A normal distribution with unit disperstion, labeled as N(0,1), is shown for comparison. In both panels, the $z$ component is shown with an arbitrary vertical displacement of $10^{-1}$.}
  \label{iner:PDF}
\end{figure}

Probability density functions, normalized by their standard deviation, are shown in Figure~\ref{iner:PDF}. Panel (a) presents the results for the four numerical simulations considered. As for the experiments both the Stokes number based on the particle radius, $St_R$, and the ratio $R/\eta$, lie between the corresponding values of simulations DNS0.2 and DNS0.5 (see Table~\ref{Iner_table}), Fig.~\ref{iner:PDF}(b) offers a comparison between these two simulations and the experiment. 

We observe a sub-Gaussian tendency in the distributions obtained from the experimental data, which is well captured by the inertial point-particles in the simulations DNS0.2 and DNS0.5 (see Fig. \ref{iner:PDF}(b)). The values of the kurtosis $\kappa$ for those datasets are $\kappa_x^\textrm{DNS0.2} = 2.62$, $\kappa_x^\textrm{DNS0.5} = 2.62$, and $\kappa_x^\textrm{EXP6} = 2.56$ for the horizontal ($x$) component; whereas for the axial ($z$) component we obtain $\kappa_z^\textrm{DNS0.2} = 2.96$, $\kappa_z^\textrm{DNS0.5} = 3.14$, and $\kappa_z^\textrm{EXP6} = 3.02$. Incidentally, the flatness values for the horizontal component of the velocity are very close to those reported in \citep{volk2011}. The effects of the large-scale anisotropy of the flow are evident in the PDFs: while the distribution for the horizontal component is sub-Gaussian, the statistics in the axial direction presents a kurtosis close to the Gaussian value. For the numerical data, it is also interesting that as the particle relaxation time is further increased (or equivalently, the Stokes numbers are increased), tails in the distributions become heavier, as is clearly seen in Fig.~\ref{iner:PDF}(a).

This behavior could be explained by a preferential sampling of the flow by the inertial particles, further enhanced by the particle increasing inertia \citep{falkovich2004, gibert2012}, or by the presence of the large-scale flow. Regarding the possible preferential sampling of some large-scale regions of the flow, we verified that in our datasets inertial particles explore more or less the same regions in the experiments as in the simulations. In the axial ($z$) direction, PDFs of particles' positions show no clear tendency towards sampling any specific region (i.e., PDFs of axial positions are approximately flat). In the horizontal ($x$) direction, particles in the experiments also show no clear preference, while particles in the simulations with $\tau_p = 0.2$ and $0.5$ show a small tendency to be near the edges of the Taylor-Green cells. However, the probability of finding particles in the third of the domain closer to the edges is only $1.1$ larger than of finding particles in the center of the cell. Moreover, this tendency is not present in the datasets with $\tau_p = 1.5$ and $\tau_p = 3.0$. In spite of these similarities in the mean exploration of the large-scale flow by the particles, the tails in the PDFs of simulations DNS1.5 and DNS3.0 (with larger Stokes numbers) deviate significantly from the data from EXP6. Thus, we conclude that these differences are due to the effective Stokes times in these simulations being too large.

\subsection{Velocity power spectra}

\begin{figure}[b]
  {\includegraphics[width=0.5\textwidth]{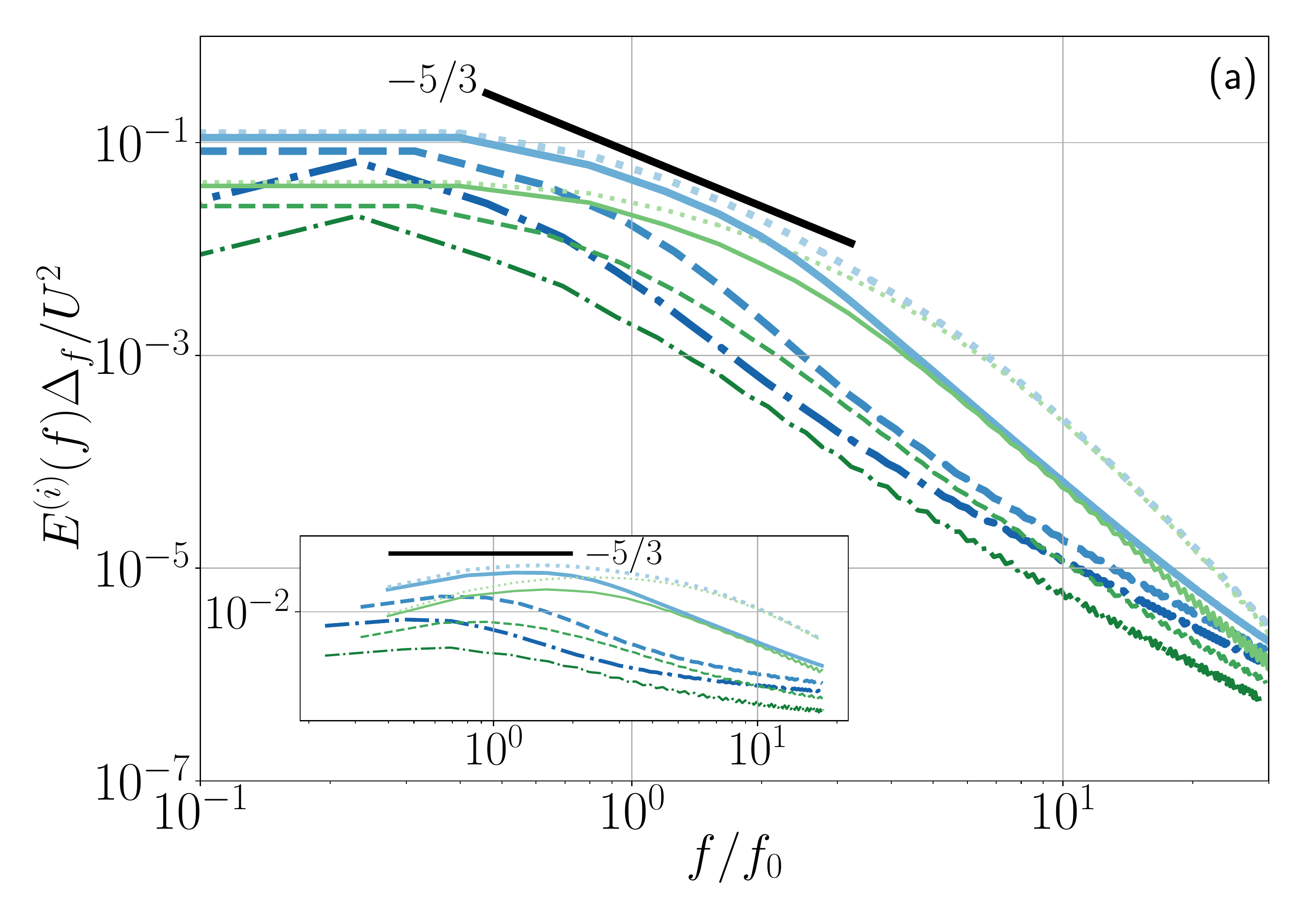}}%
  \hfill%
  {\includegraphics[width=0.5\textwidth]{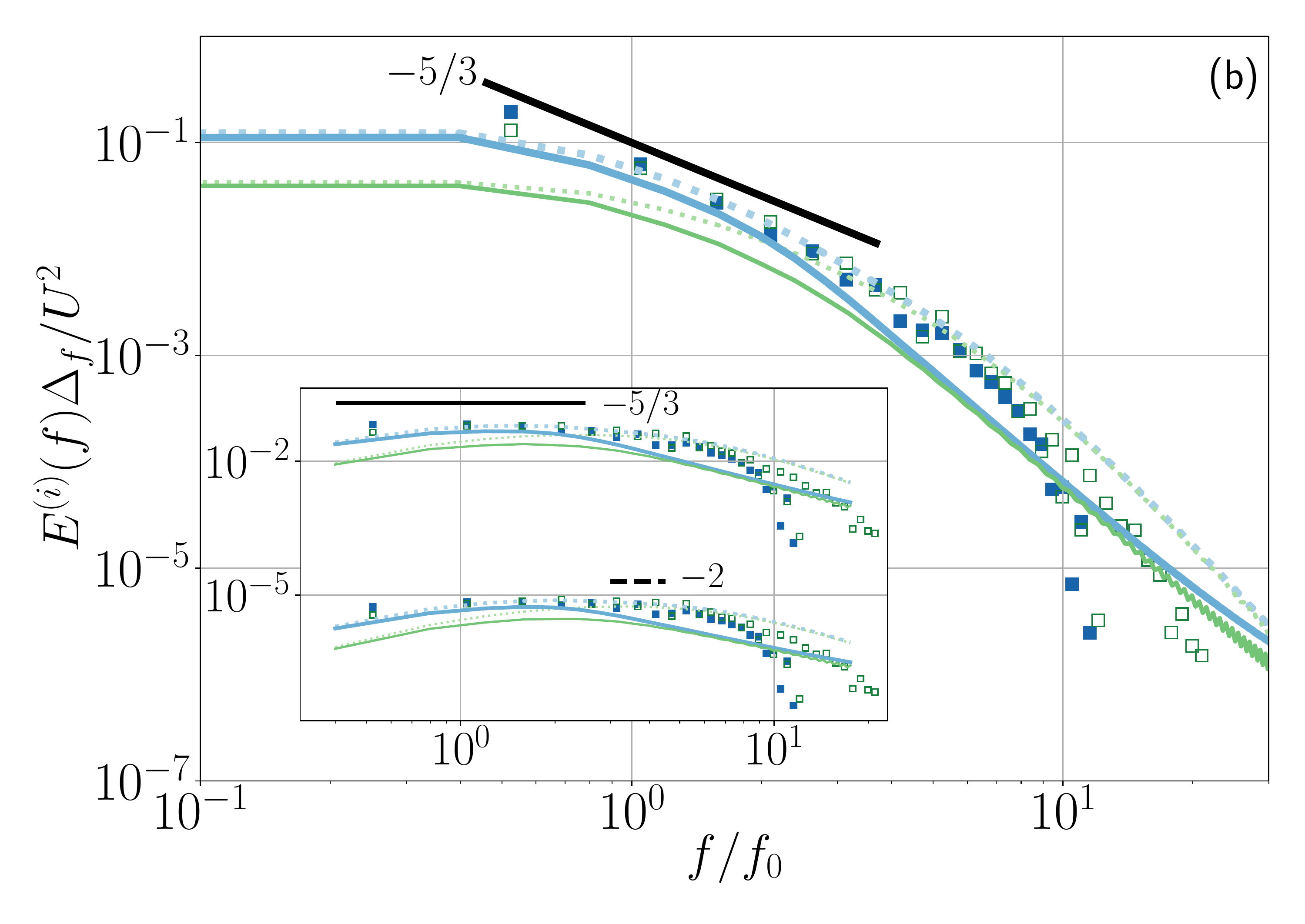}}
  \caption{Inertial particles' velocity power spectra in log-log scale. Labels are the same as in Fig. \ref{iner:PDF}. (a) Numerical results, for particles with different response time $\tau_p$. The inset shows the spectra compensated by a power law with exponent $-5/3$. (b) Comparison of the experimental spectrum EXP6 with numerical simulations DNS0.2 and DNS0.5. The inset shows the spectra compensated by $-5/3$ (above) and $-2$ power law exponents (below). The $-2$ compensated spectra was shifted vertically by a factor $10^{-4}$ for clarity.}
  \label{iner:energy_spec}
\end{figure}

The inertial particles' velocity power spectrum was computed for the $x$ and $z$ components of the velocity. The results for the DNSs are shown in Fig.~\ref{iner:energy_spec}(a). A power law compatible with a $-5/3$ exponent is observed in the data corresponding to DNS0.2 and DNS0.5 in a short frequency range. Moreover, as $\tau_p$ increases further in the simulations, almost no power law can be discerned; still, in the spectra compensated by $-5/3$ (see the insets in Fig.~\ref{iner:energy_spec}) a narrow plateau may still be identified at low frequencies in all datasets. The narrower $-5/3$ scaling range might be linked to the fact that as the particle response time grows, the particle becomes less sensitive to fluctuations in frequencies $f\gg 1/\tau_p$. To illustrate this we can consider fields in Eq.~\eqref{eq:iner} as random variables, and Fourier transform this equation. Taking the absolute value results in
\begin{equation}
    |\hat{{\bf v}}_f|^2 = \frac{|\hat{{\bf u}}_f|^2}{1 + (2\pi f\, \tau_p)^2},
    \label{eq:iner_spec}
\end{equation}
where $\hat{{\bf v}}_f$ and $\hat{{\bf u}}_f$ are respectively the Fourier transforms of ${\bf v}$ and ${\bf u}$. Note this equation can be interpreted as a filter: fluctuations in the fluid velocity ${\bf u}$ (at the particle position) with frequency $f\gg 1/\tau_p$ are attenuated in ${\bf v}$. Moreover, assuming that the particle samples the flow in the same way as a tracer (which is not entirely correct, as preferential flow sampling has been observed for particles with inertia and finite-size \citep{monchaux2012}), then ${\bf u}({\bf x}_p, t)$ should have the same spectral properties as the power spectrum of the tracer's velocity. Under these hypothesis, ${\bf v}$ should have a power spectrum similar to the spectrum in Fig.~\ref{lag:energy_spec} at low frequencies, and damped amplitudes for large frequences (compared with $1/\tau_p$). Increasing $\tau_p$ should also result in a stronger filter acting at smaller frequencies. Even though these arguments neglect the effect of (small-scale) preferential sampling, the conclusions are in qualitative agreement with the observed spectra.

Figure~\ref{iner:energy_spec}(b) exhibits the experimental data compared with the data from DNS0.2 and DNS0.5. The experimental data presents a power law compatible with $-5/3$ for almost a decade of frequencies. The $-2$ power law, which was present in the tracers measurements, appears here only for a very short range. The particle's finite-size effectively filters the fluctuations at intermediate frequencies corresponding to the inertial range. Consequently, the $-2$ range shortens significantly when compared with the tracers. Both at low and at intermediate frequencies (up to $f/f_0\approx 8$) the DNSs and the experimental data are in good agreement (slightly better when EXP6 is compared with DNS0.2). This again indicates, as with the PDFs of the particles' velocities, that numerical simulations of point particles in a Taylor-Green flow can mimic some statistical properties observed in the von K\'arm\'an periments when the Stokes number is estimated using the particles' effective response time $\tau_R$.

Moreover, note that even though the numerical datasets DNS0.2 and DNS0.5 have similar $St_R$ and $R/\eta$, the ``filtering" of the spectrum observed in the experiment is sharper: while the DNS and experimental spectra coincide up to $f/f_0\approx 8$, for larger frequencies the experimental spectrum decays much faster than in the simulations. This is an indication that to fully capture the behavior of the particle in the experiments other effects need to be considered in the simulations, such as buoyancy, added mass effects, or other effects related to the finite size of the particles \cite{maxey1983}. However, the behavior observed in the experiments can be also mimicked if the filtering of the fluid velocity by the particle is of the form
\begin{equation}
    |\hat{{\bf v}}_f|^2 = \frac{|\hat{{\bf u}}_f|^2}{1 + (2\pi f\, \tau_p )^\alpha},
    \label{eq:spec_exp}
\end{equation}
where $\alpha$ controls how abruptly the inertial particle power spectrum deviates from the fluid velocity power spectrum. As $\alpha$ grows, the spectrum $|\hat{{\bf v}}_f|^2$ retains the behavior of $|\hat{{\bf u}}_f|^2$ better up to $f\approx 1/\tau_p$, and decays faster for $f\gg 1/\tau_p$. Note the model in Eq.~\eqref{eq:iner} can be also modified to result in such a sharper decay.

\subsection{Velocity autocorrelation functions}

\begin{figure}
  {\includegraphics[width=1\textwidth]{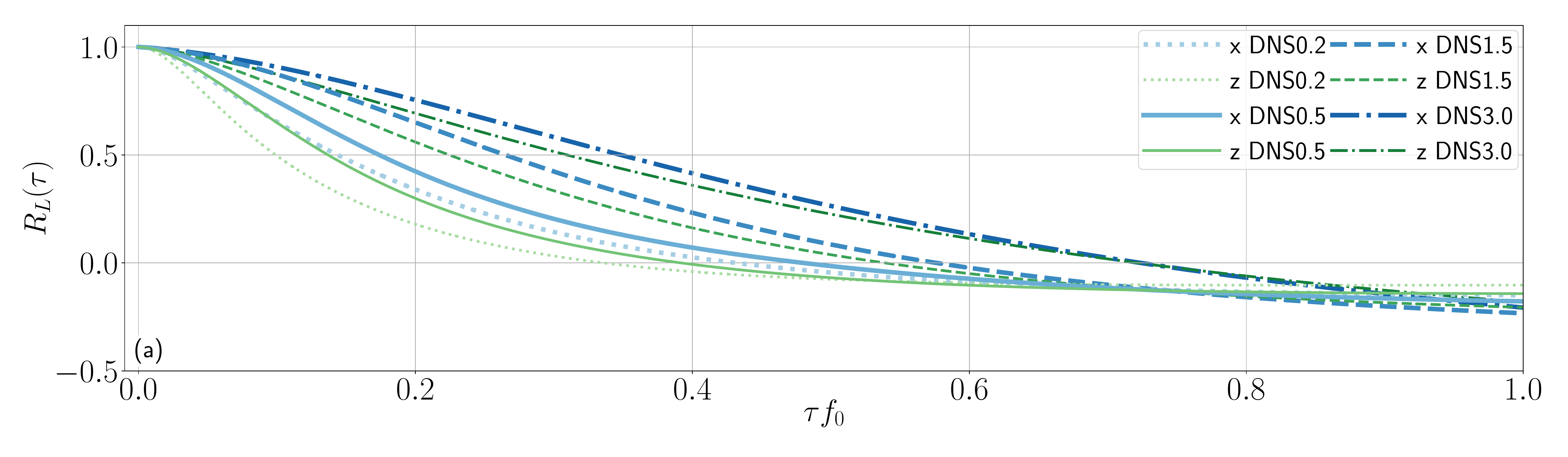}}%
  \hfill%
  {\includegraphics[width=1\textwidth]{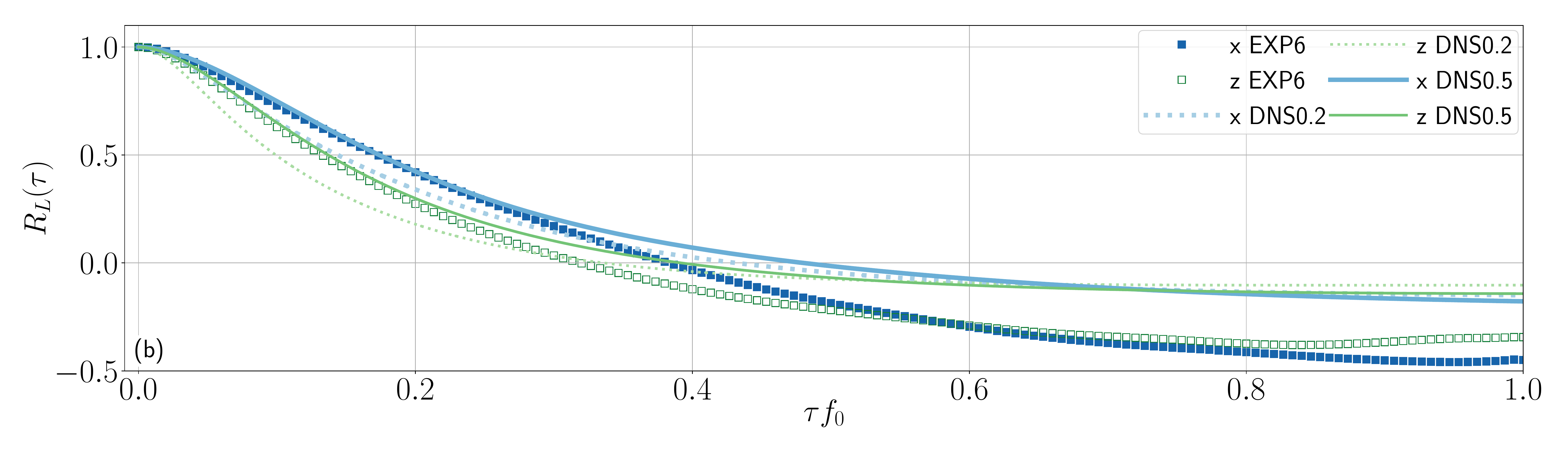}}
  \caption{Inertial particle velocity autocorrelation functions for the horizontal ($x$) and axial ($z$) components of the velocity in both experiments and simulations. The time axis is normalized by $f_0$, the rotation frequency of the blades in the experiment, and the frequency associated to the largest eddy turnover time in the DNS. Panel (a) corresponds to the four numerical runs, while panel (b) shows the experimental data alongside DNS0.2 and DNS0.5.}
  \label{iner:corr}
\end{figure}

The inertial particles' velocity autocorrelation function is computed using the definition in Eq.~\eqref{eq:corr}. The curves for the DNSs are shown in Fig.~\ref{iner:corr}(a). It can be seen that the larger $\tau_p$, the smaller the anisotropy between the $x$ and $z$ coordinates: the zero-crossing times of the two velocity components become more similar (see values in Table~\ref{Iner_table}). At the same time, as $\tau_p$ increases, the particles decorrelate more slowly. This is in agreement with the simple ``filter" model discussed in the previous section. Since the velocity autocorrelation function is related to the energy spectrum via the Wiener-Khinchin theorem, the Fourier transform of the expression in Eq.~\eqref{eq:iner_spec} is the velocity autocorrelation function $C_v(\tau)$ associated to that spectrum. By numerical computation of the Fourier transform of Eq.~\eqref{eq:iner_spec} (using the spectrum $|\hat{{\bf u}}_f|^2$ obtained from the tracers), we can confirm that the autocorrelation decays faster for smaller values of $\tau_p$, which is consistent with the behavior observed in the data.

In Fig.~\ref{iner:corr}(b) the experimental data is compared with the numerical datasets DNS0.2 and DNS0.5. A similar decay is observed in the EXP6 and DNS0.5 curves, specially for time lags $\tau\, f_0 \leq 0.3$. The zero-crossing times $\tau_L^{(i)}$ (for $i=x$ or $z$) are again comparable between the experiment and simulations DNS0.2 and DNS0.5 (see also Table~\ref{Iner_table}). An even better agreement is obtained if instead of $\tau_L^{(i)}$, we compute a component-wise ``particle integral time" as
\begin{equation}
T^{(i)} = \int_0^{\tau_{95}} R_L^{(i)}(\tau) \: d\tau ,   
\end{equation}
where the time $\tau_{95}$ is the time needed for $R_L^{(i)}(\tau)$ to decrease by 95\%. This definition was introduced by \citet{machicoane2016} to better quantify decorrelation times of inertial particles in von K\'arm\'an experiments; here we use the same definition but, following our motivation to compare bulk flows, we do not apply any specific method to try to reduce the effects of mean flow contributions. The values for this time are also shown in Table~\ref{Iner_table}; we find that the product $T^{(x)} f_0$, as well as the ratio $T^{(x)}/T^{(x)}$, are similar for EXP6, DNS0.5, and DNS0.2, accounting for the similar decay of $R_L(\tau)$, and confirming once again certain statistical agreement between the behavior of inertial particles in the von K\'arm\'an experiment, and of the modeled particles in the Taylor-Green simulation.

\subsection{Structure functions\label{sec:inerstuct}}

\begin{figure}
  {\includegraphics[width=0.5\textwidth]{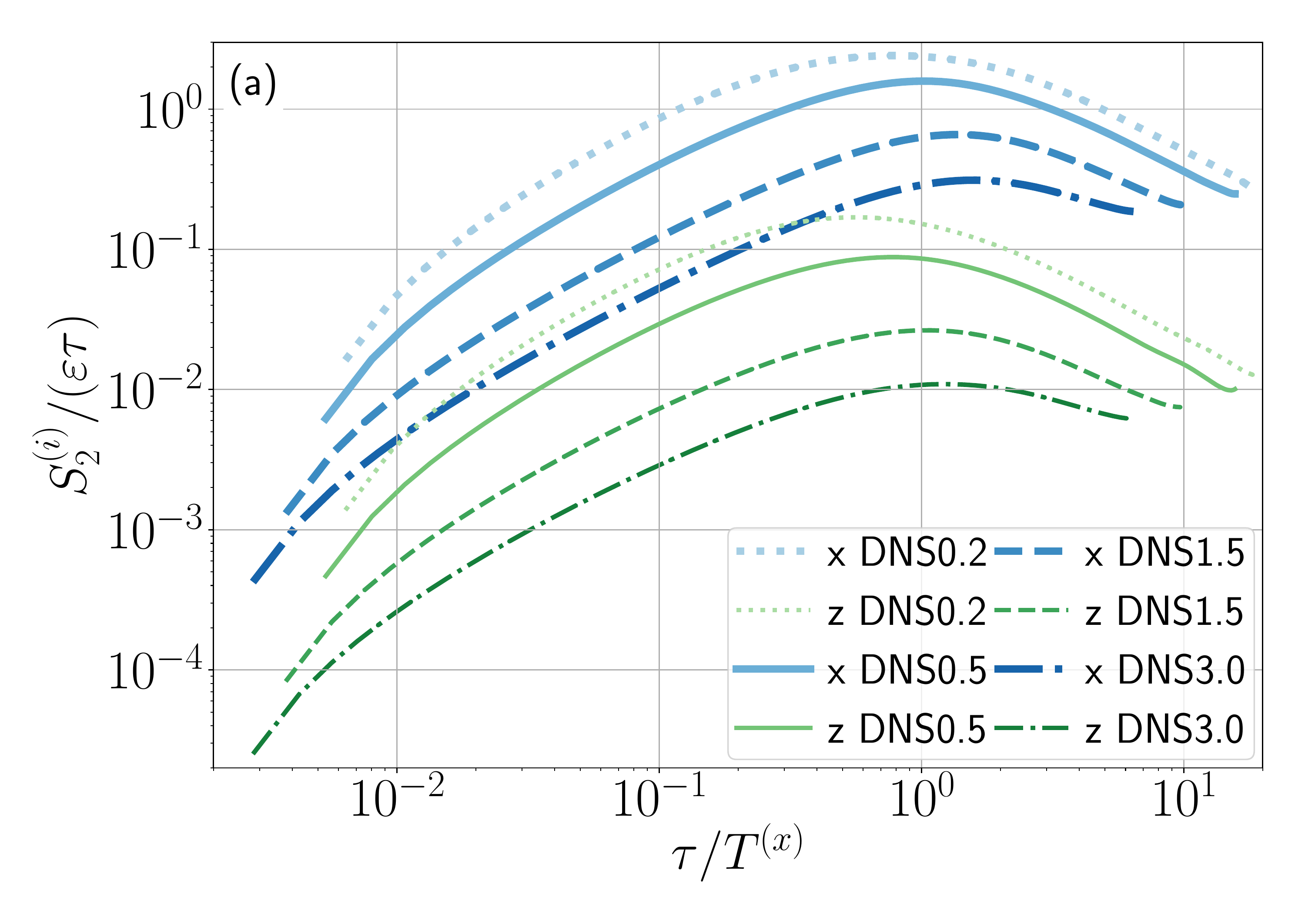}}%
  \hfill%
  {\includegraphics[width=0.5\textwidth]{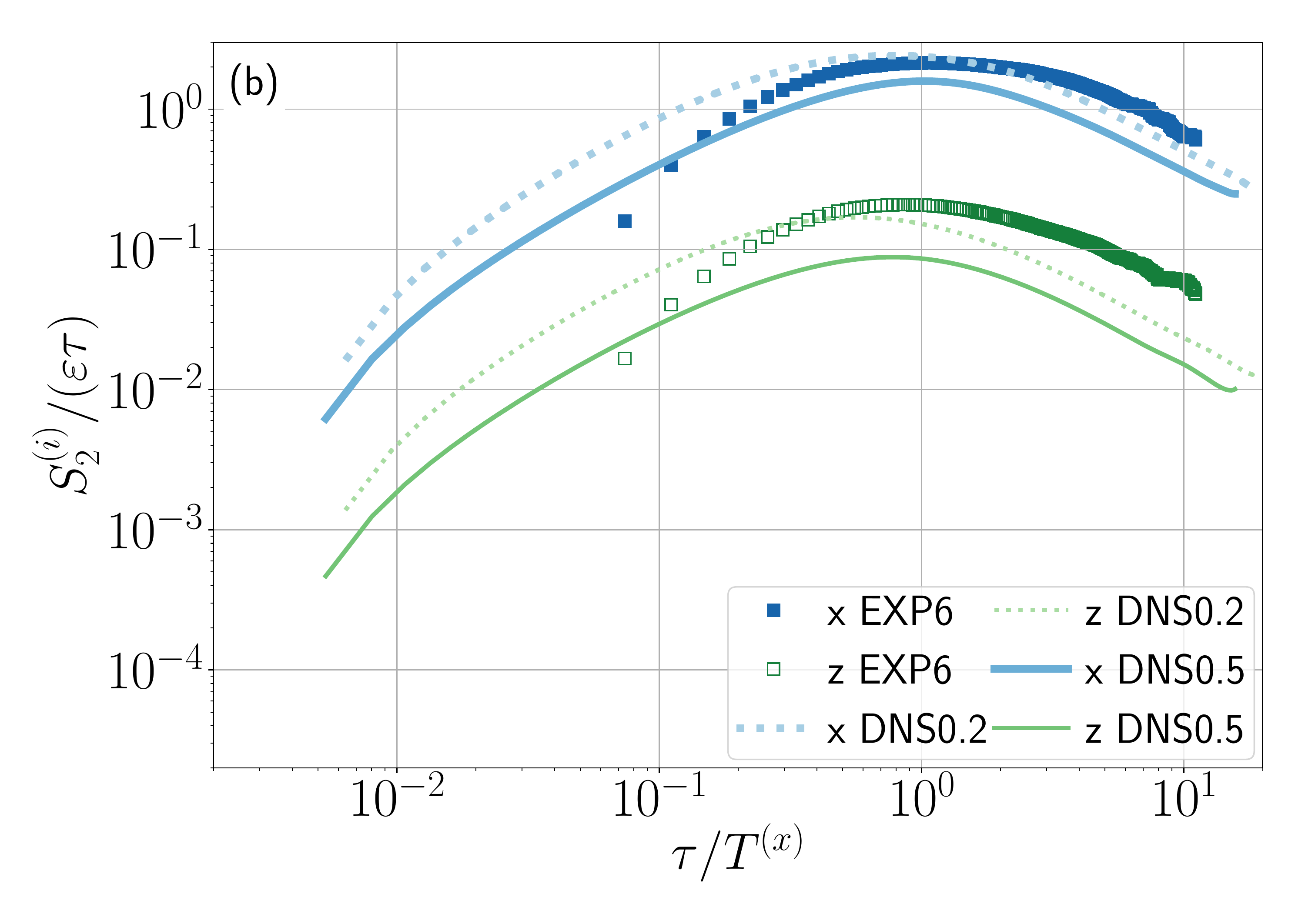}}
  \caption{Inertial particles' velocity second order structure function, compensated by $\varepsilon\tau$, the prediction for the Lagrangian inertial range of the fluid. Time lags in the figures are made dimensionless using the particle inertial time $T^{(x)}$. Panel (a) shows the data from the DNSs, for the Cartesian velocity components $x$ and $z$, while panel (b) compares the experimental data with DNS0.2 and DNS0.5. In both figures the $z$-component curves are shown with an arbitrary vertical displacement of $10^{-1}$ for clarity.}
  \label{iner:S2}
\end{figure}

The one-dimensional second order velocity structure function is calculated as in Eq.~\eqref{eq:S2} for the inertial particles. The numerical data, compensated by the Lagrangian prediction for the inertial range, is shown in Fig.~\ref{iner:S2}(a). No plateau is present, and the amplitude of the curves decreases with increasing $\tau_p$. The absence of a plateau (which is already barely visible in tracers' measurements when trajectories in all the cell are considered), is in this case also a consequence of the particles' insensitivity to fluctuations in time scales $\tau < \tau_p$, combined with the slow convergence of $S_2(\tau)$ towards its asymptotic value in the inertial range discussed in Sec.~\ref{sect:lag-tracers}.
On the other hand, as $S_2(\tau) = 2(C_v(0)) - C_v(\tau))$, the numerical estimation of $S_2(\tau)$ from Eq.~\eqref{eq:iner_spec} indicates that increasing $\tau_p$ has the same effect observed in our data: the amplitude of the curves decreases as $\tau_p$ grows. Since $S_2(\tau)$ is a measure of how disperse the particle's velocity increments are for a given time lag $\tau$, a decrease in its amplitude can be also thought of as a smoothing of the velocity signal.

In Fig.~\ref{iner:S2}(b), the compensated data from EXP6 is compared with the numerical data from DNS0.2 and DNS0.5. The overall shape of the curves is similar, notably they reach their maximum value at nearly the same value for the ratio $\tau/T^{(x)}$. Note in this case the time axis is normalized by $T^{(x)}$ instead of using $\tau_L^{(x)}$, as it was shown that the former time captures better the similarities in the decay of the velocity autocorrelation functions. The experimental data, as in the case of the tracers, displays a plateau for a narrow range of frequencies, and is less anisotropic than that obtained from the simulations.

\subsection{Statistics of particles' velocity fluctuations in subregions}

\begin{figure}
  {\includegraphics[width=0.5\textwidth]{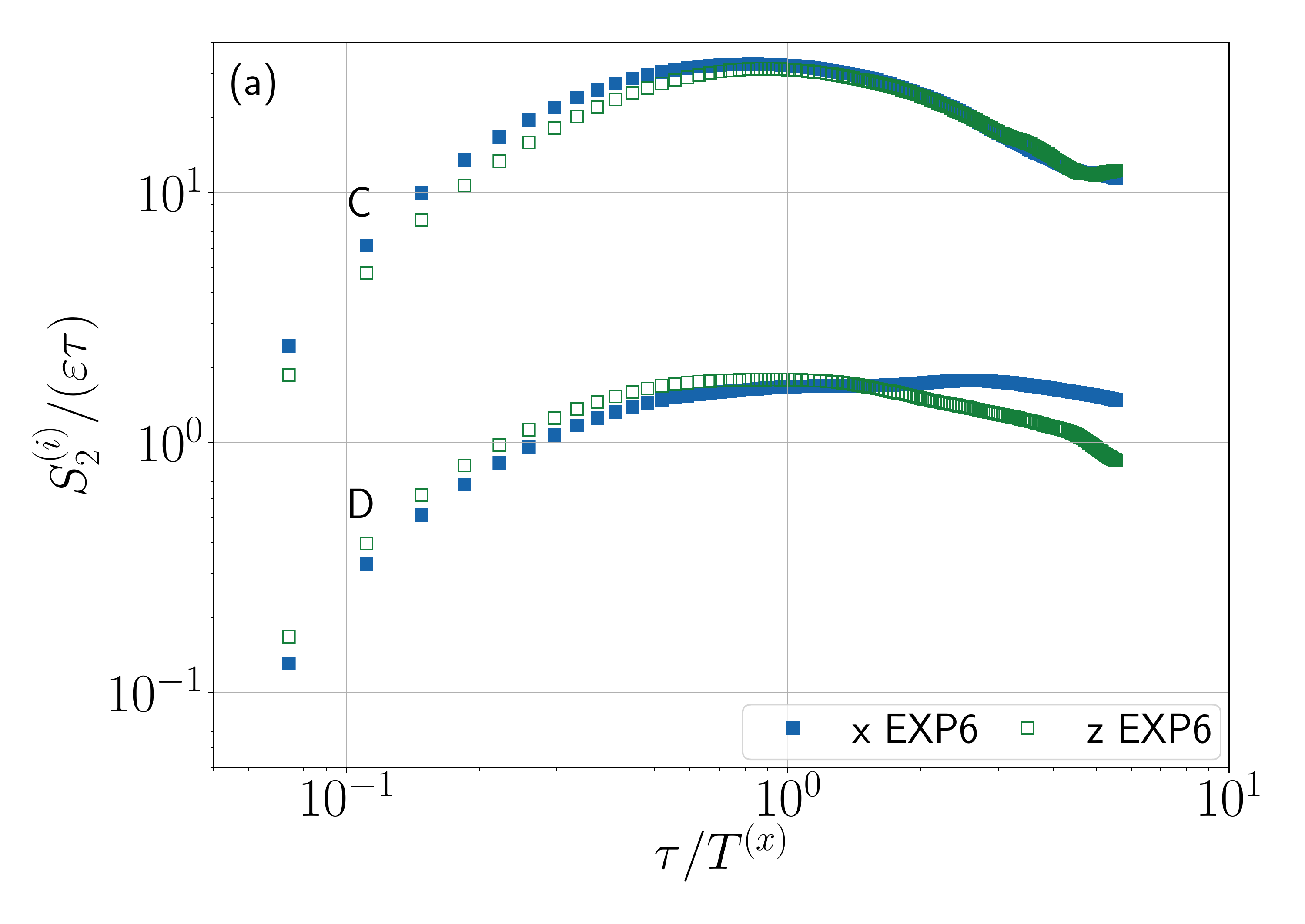}}%
  \hfill%
  {\includegraphics[width=0.5\textwidth]{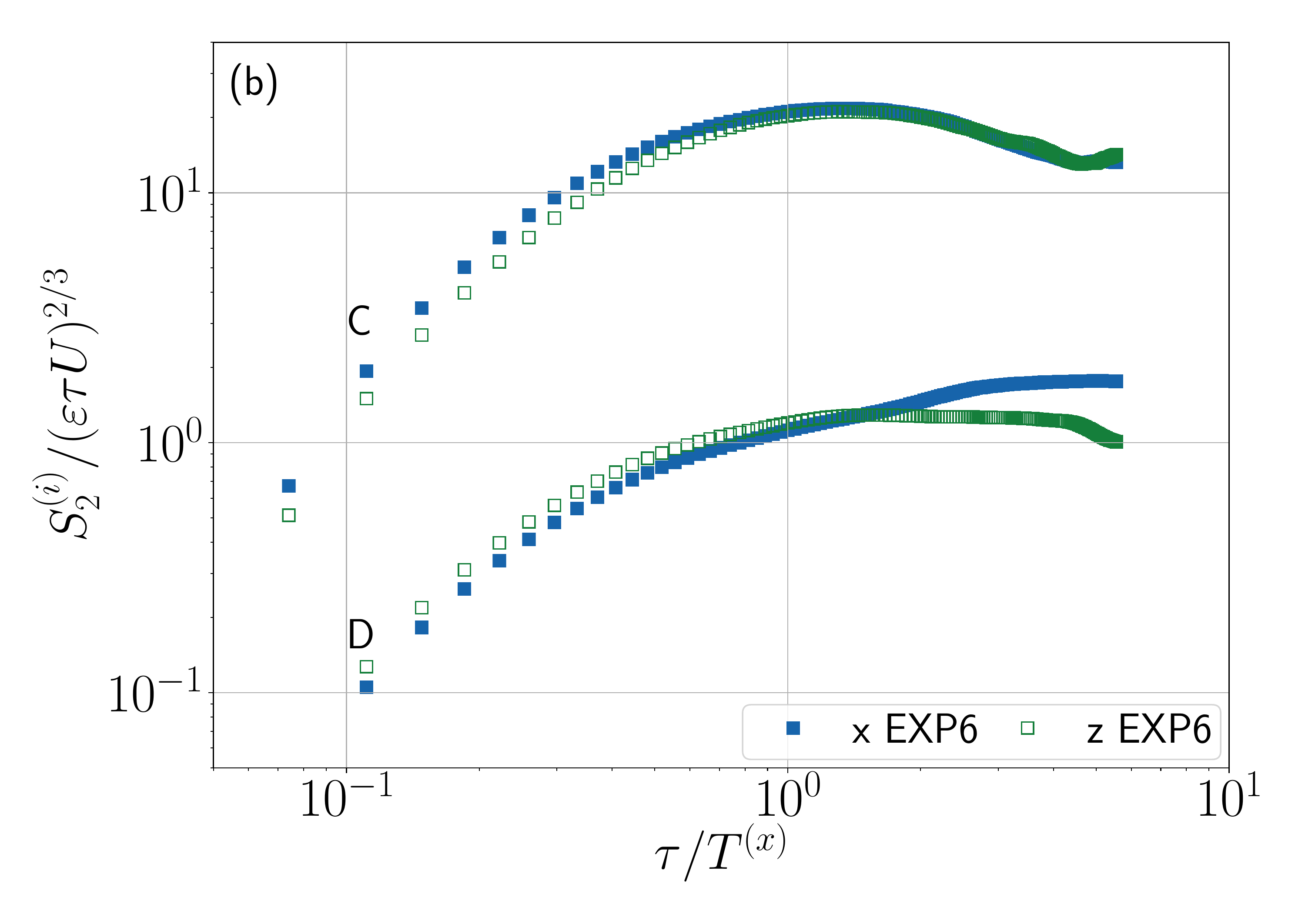}}
  \caption{Experimental inertial particle's second order velocity structure functions, compensated by $\varepsilon \tau$ in panel (a) for trajectories conditioned to subregion C (top curves, displayed for clarity with a vertical displacement by multiplying them by a factor of $10$), and to subregion D (bottom curves). Panel (b) shows $S_2(\tau)$ compensated by $(\varepsilon \tau U)^{2/3}$, with the same convention as in panel (a) for trajectories conditioned to subregions C and D. Numerical results are not shown for clarity, but display the same qualitative behavior.}
  \label{iner:S2_exp_z}
\end{figure}

As in the case of tracers, the second order velocity structure function is computed conditioning the inertial particles' trajectories to the same two regions considered in Sec.~\ref{sec:lag_subregions}, i.e., near the center of the cell (subregion C) and near the disks (subregion D). The data corresponding to EXP6 for the two subregions is displayed in Fig.~\ref{iner:S2_exp_z}(a) compensated by $\varepsilon\tau$, and in Fig.~\ref{iner:S2_exp_z}(b) compensated by $(\varepsilon\tau U)^{2/3}$. The curves conditioned to subregion C show no considerable difference between the two Cartesian components $x$ and $z$. At the same time,  a short plateau appears at $\tau/T^{(x)} \lesssim 1$ for $S_2/(\varepsilon\tau)$, although narrower than the plateau observed for the tracers in the same region of the cell. A different behavior is observed for the data conditioned to subregion D. The curves display a clear anisotropy for time lags $\tau \geq T^{(x)}$, the time scales in which the mean flow can be expected to become dominant. The curves for the axial component of the velocity behave similarly than for the tracers', with nearly half a decade compatible with $S_2 \propto \tau^{2/3}$. The horizontal velocity component, however, shows no plateau on this range of time scales when compensating $S_2$ by $\tau^{2/3}$. This may be linked to the effect of the energy injection mechanism on the particles' trajectories close to the disks. Comparison with the simulations is qualitatively the same as in Fig.~\ref{iner:S2}, and not shown in this figure for clarity.

\subsection{Acceleration spectra}

\begin{figure}
  {\includegraphics[width=0.5\textwidth]{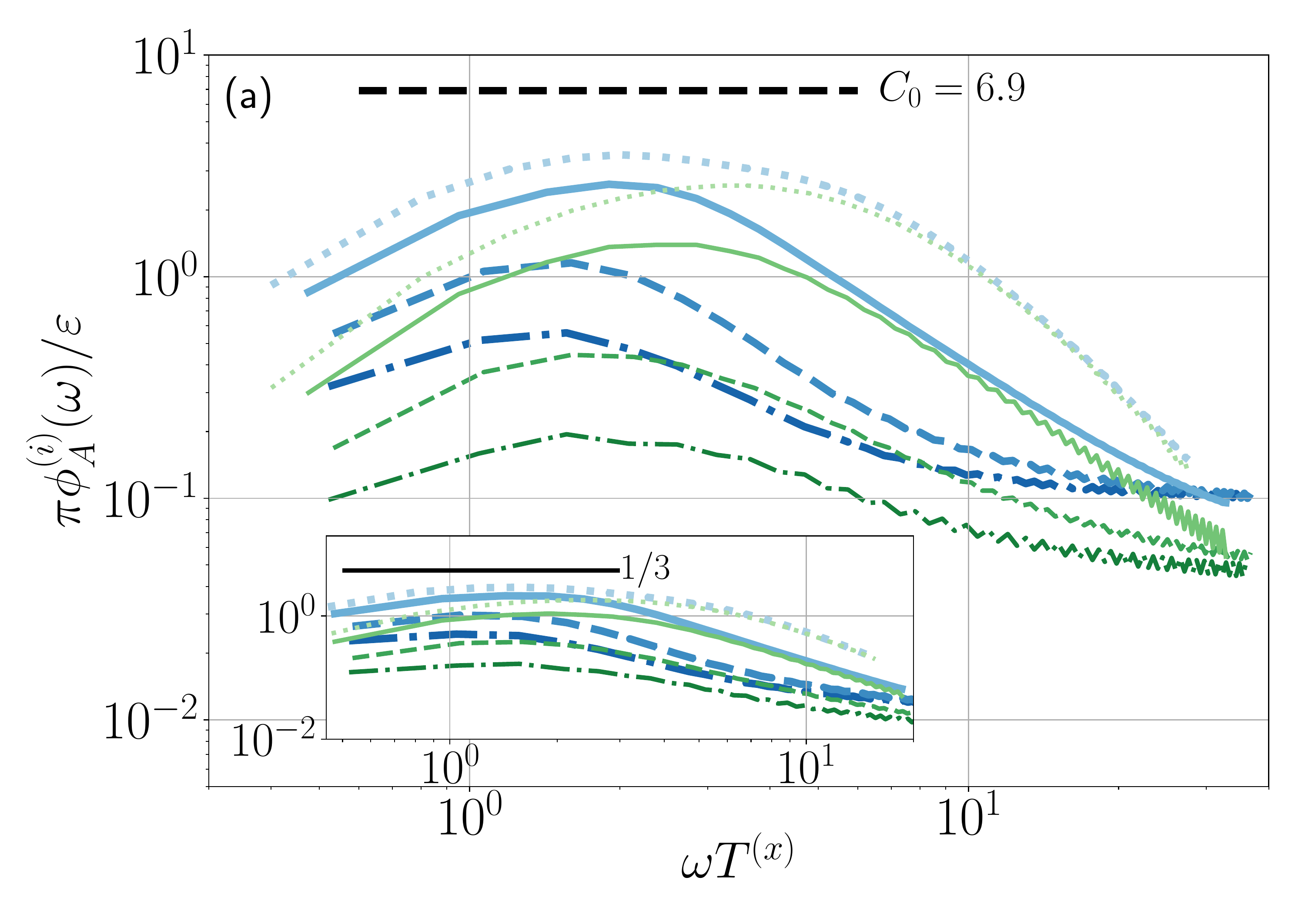}}%
  \hfill%
  {\includegraphics[width=0.5\textwidth]{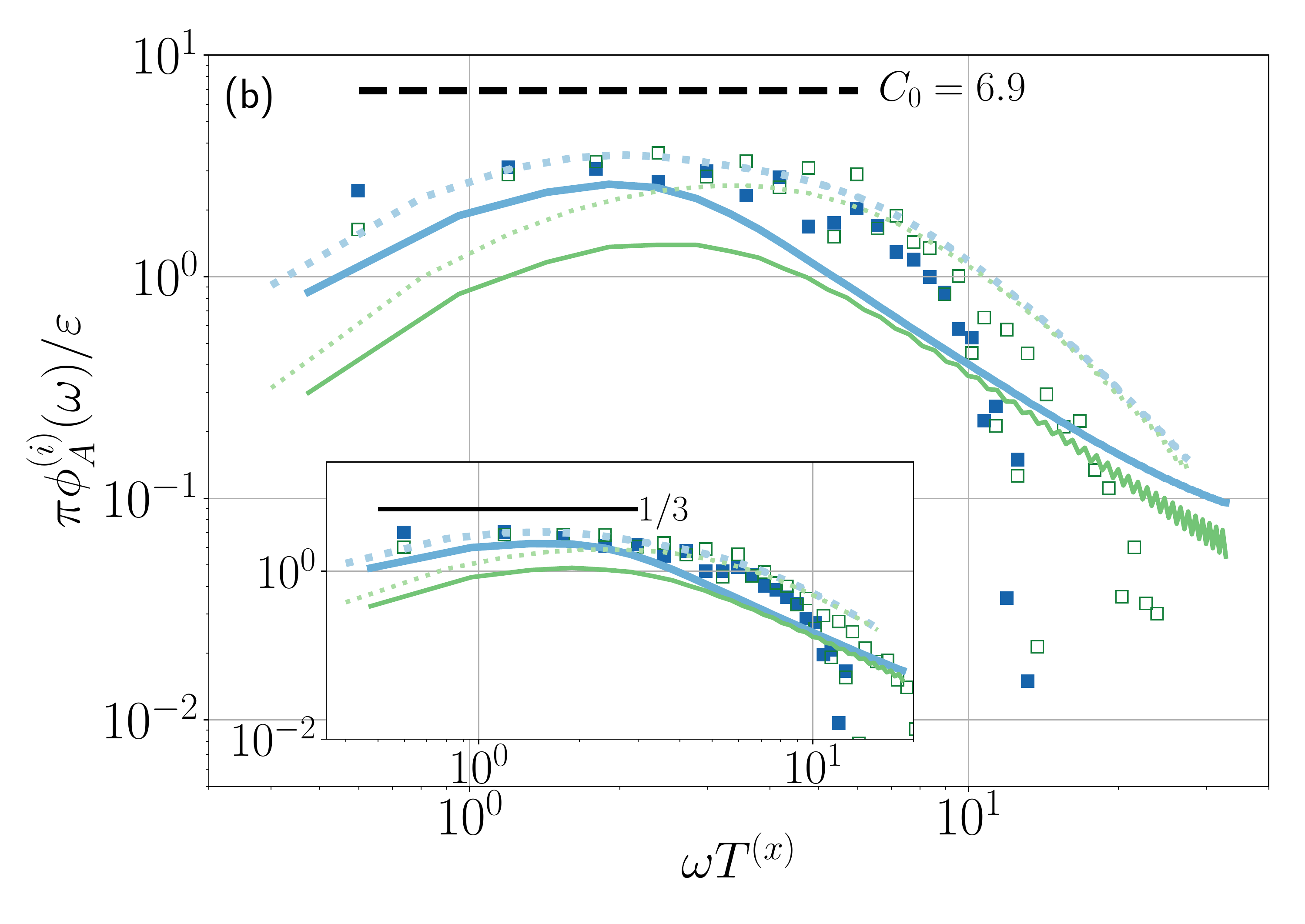}}
  \caption{Inertial particles' acceleration spectra normalized by the energy injection rate $\varepsilon$, in panel (a) for all the DNSs, and in panel (b) for experiment EXP6 and for simulations DNS0.2 and DNS0.5. References for the curves are the same as in Fig.~\ref {iner:S2}. The horizontal dashed line with $C_0 = 6.9$ corresponds to the asymptotic amplitude expected in the inertial range for Lagrangian trances, and is only given as a reference. The insets show the spectra compensated by a power law with exponent $1/3$.}
  \label{iner:acc_spec}
\end{figure}

Finally, the inertial particles' acceleration spectrum is computed from the velocity autocorrelation function using the definition in Eq.~\eqref{eq:acc_spec}. The spectra, normalized as for the case of the tracers, is plotted in Fig.~\ref{iner:acc_spec}(a) for the DNSs. For DNS0.2, the numerical dataset with the smallest value of $\tau_p$, a narrow plateau may be present in the vicinity of $\omega T^{(x)} \approx 3$. No plateau is observed in the rest of the curves, which is consistent with the absence of clear scaling laws in the velocity power spectrum of the inertial particles. At the same time, the amplitude of the curves decreases as $\tau_p$ grows. Since we can estimate $\sqrt{\langle a^2(\tau)\rangle} \sim \langle v(t+\tau) - v(t)\rangle_t /\tau$, a smaller amplitude of the acceleration spectrum is compatible with a smaller $\sigma_{v\tau}$, the dispersion of the velocity differences at a given $\tau$, which was already observed to decrease with increasing $\tau_p$ in the second order structure functions in Sec.~\ref{sec:inerstuct}. An inset with the acceleration spectra compensated by $\omega^{-1/3}$ (the sweeping-dominated scaling) is also shown as a reference in this figure; a narrow plateau can be identified for the smaller frequencies in the vicinity of $\omega\, T^{(x)} \approx 10^0$.

In Fig.~\ref{iner:acc_spec}(b) the experimental data is compared with DNS0.2 and DNS0.5. Even though the DNS data is more anisotropic, the amplitude and overall shape of the experimental spectra are similar to those of DNS0.5. However, the experimental curves do display a plateau for almost a decade of frequencies, but with an amplitude smaller than that of the tracers. This difference between the simulations and the experiments further suggests that additional effects need to be considered in the simulations in order to achieve a more detailed description of the behavior of the particle dynamics in the experiments. Still, all the datasets present a short plateau when compensated by $\omega^{-1/3}$ (see inset in Fig. \ref{iner:acc_spec}(b)), consistently with the $-5/3$ range observed in the velocity power spectra in Fig.~\ref{iner:energy_spec}(b).

\section{Discussion}

In this work we presented a comparison of tracers' and finite-size inertial particles' velocity and acceleration statistics in two paradigmatic turbulent swirling flows: an experimental von K\'arm\'an flow, and a numerical flow obtained by imposing a Taylor-Green mechanical forcing. For the simulations of the inertial particles, a simple point particle model was used, which was considered as an effective model with an effective particle response time.

In spite of the differences in boundary conditions and in the forcing mechanisms, scaling and statistical properties of tracers share similarities between both flows, and also display a clear effect of the mean flow on particle dynamics which in turn affect turbulent statistical properties. Results from the experiments and the simulations show good agreement in the decorrelation times of the tracers' velocity, and also in the power laws observed in the velocity power spectrum. We find two different power law exponents: a range compatible with $-2$ for sufficiently small frequencies, and one compatible with $-5/3$ for frequencies associated to the large-scale motion of the flow. In both the numerical simulations and the experiments, the compensated tracers' velocity second order structure function has a short plateau corresponding to the Lagrangian inertial range. A more clear scaling range is also seen, associated to the effect of the mean flow. Similar results are obtained in the experiments and the simulations when the acceleration spectrum is considered, up to high frequencies that can be associated to the dissipative range. Remarkably, our results for the second order Lagrangian structure functions and the acceleration spectra indicate that sweeping by the large-scale flow plays a relevant role in the particle evolution even in the Lagrangian frame. This is further confirmed by studying the statistics of tracers' velocities in subregions close to the central shear layer, and in subregions close to the forcing mechanism (either the disks in the experiments, or the maximal Taylor-Green forcing in the simulations). Statistics conditioned to the former subregion show a clearer Lagrangian inertial range scaling, while statistics conditioned to the latter subregion indicate stronger sweeping effects.

Having found good agreement between the experiment and the numerical simulations from the tracer's dynamics viewpoint, we compare the velocity and acceleration statistics of an inertial particle. We find that a suitable way to compare finite-size particles in the experiments with inertial point-particles in the numerics is via an effective Stokes number based on the eddy turnover time at the particle radius. Even though in our simulations only viscous drag is considered, several statistical quantities in the experiments are well captured by the simulations; these include the behavior of the velocity probability distribution function, the velocity power spectrum, the amplitude of the second order velocity structure function, and the acceleration spectrum. However, the ``filtering" of fast fluctuations in the flow by the inertial particles is more accentuated in the experiment than in the DNSs. Such a deficiency can be corrected by a simple modification to the equation for the evolution of the particles. This also points out that important physical effects are missing in the simple phenomenological model we consider for our inertial particles. Other effects were taken into account and shown to be relevant in Ref.~\citep{Volk2008}, although still under the assumption of small particles. Indeed, most models for inertial particles are imperfect in this sense. Our results indicate that not only other forces acting over the particles should be taken into account, but that also the impact of the large scale flow on the particles must be considered in the comparisons.

The comparative analysis presented here between experimental data and numerical simulations also led us to introduce definitions of dimensionless numbers that allow comparisons between the von K\'arrm\'an and Taylor-Green flows. In particular, definitions of Reynolds numbers, Taylor-based Reynolds numbers, dimensionless autocorrelation times, and Stokes numbers were provided that are in good agreement between the experiments and the simulations that display similar behavior for the different particles studied.

Finally, in spite of the similarities in the results, there are important discrepancies and limitations that are worth mentioning, and that open new paths for future studies. Firstly, only one inertial particle was used in the experiments, and a detailed exploration of particles with different masses and radius is needed to better calibrate effective dimensionless numbers for the simulations. Secondly, the large-scale flow in the simulations displays a stronger anisotropy in its velocity components than the flow in the experiments. To improve comparisons between experiments and simulations, the geometry of the blades in the impellers could be changed, to change the ratio of the vertical to horizontal velocities in the von K\'arm\'an flow (as was done before, e.g., in von K\'arm\'an dynamo experiments, see \cite{monchaux2009}). And thirdly, our numerical model for the particles is based on a model for point particles that only considers the Stokes drag, and as a result our particles response times can only be interpreted in an effective way. Other effects, such as added mass effects, buoyancy, or finite size effects such as the Basset-Boussinesq history term or the Fax\'en corrections \cite{maxey1983} should be taken into account to improve particle modeling, and to properly consider the spatial variation of the flow in the vicinity of the finite-size particles.

\begin{acknowledgments}
The authors acknowledge financial support from grants UBACYT No.~20020170100508BA and PICT No.~2015-3530.
\end{acknowledgments}

\bibliography{ms}

\end{document}